\documentclass[11pt]{article}
\pdfoutput=1
\usepackage{graphicx}
\usepackage{hyperref}
\usepackage{epsfig}
\usepackage{amsmath}
\usepackage{amsthm}
\usepackage{amssymb}
\usepackage{multirow}
\usepackage{color}
\usepackage[nosort]{cite}
\usepackage{hyperref}
\usepackage[fontsize=11.45pt]{scrextend}
\usepackage{braket,mathrsfs,slashed}
\usepackage{float}
\usepackage{setspace}
\usepackage[caption=false]{subfig}
\usepackage{comment}
\usepackage{bm}
\newlength{\abstractwidth}
\newcommand{\be}{\begin{equation}}
\newcommand{\ee}{\end{equation}}

\renewcommand{\title}[1]{\vbox{\center\bf{\Large{#1}}}\vspace{5mm}}
\renewcommand{\author}[1]{\vbox{\center#1}\vspace{5mm}}
\newcommand{\address}[1]{\vbox{\center\em#1}}

\usepackage{dsfont}

\usepackage[utf8]{inputenc}
\usepackage{rotating}
\usepackage{tikz-feynman}

\usepackage{dsfont}
\usepackage[paper=a4paper,margin=1.5cm]{geometry}
\usepackage[utf8]{inputenc}
\usepackage{rotating}
\usepackage{soul}

\usepackage{mathrsfs} 
\usepackage{bbm}
\usepackage{etoolbox} 
\usepackage{multirow}
\renewcommand\[{\begin{equation}}
\renewcommand\]{\end{equation}}

\newcommand{\ba}{\begin{eqnarray}}
\newcommand{\ea}{\end{eqnarray}}

\newcommand{\scat}{\mathcal{M}}
\newcommand{\re}{\mathrm{Re}}
\newcommand{\im}{\mathrm{Im}}
\newcommand{\der}{\partial}

\newcommand{\coloneqq}{:=}

\hypersetup{pdfstartview=FitV,colorlinks=true,linkcolor=midblue,citecolor=midblue,filecolor=midblue,urlcolor=midblue}

\definecolor{midblue}{rgb}{0,0,0.5}

\begin{document}
	
		\newgeometry{top=3.1cm,bottom=3.1cm,right=2.4cm,left=2.4cm}
		
	\begin{titlepage}

\begin{flushright} {\footnotesize NORDITA 2023-028, CTPU-PTC-23-22}  \end{flushright}
		\begin{center}
			\hfill \\
			\hfill \\
			\vskip 0.5cm

    \title{\Large New lower bounds on scattering amplitudes: \\[1.85mm]non-locality constraints}

			\author{\large Luca Buoninfante$^{a,\,\star}$, Junsei Tokuda$^{b,\,\dagger}$ and Masahide Yamaguchi$^{c,d\,\ddagger}$
			 }
			
			\address{$^a$Nordita, KTH Royal Institute of Technology and Stockholm University\\
				Hannes Alfvéns väg 12, SE-106 91 Stockholm, Sweden	\\[1.5mm]
				$^b$Particle Theory and Cosmology Group, Center for Theoretical Physics of the Universe,\\ Institute for Basic Science (IBS), Daejeon, 34126, Korea\\[1.5mm]
                $^c$Cosmology, Gravity and Astroparticle Physics Group,\\ Center for Theoretical Physics of the Universe, Institute for Basic Science (IBS),\\ Daejeon, 34126, Korea\\[1.5mm]
				$^d$Department of Physics, Tokyo Institute of Technology, Tokyo 152-8551, Japan}
				\vspace{.3cm}

		\end{center}

\begin{abstract}
Under reasonable working assumptions including the polynomial boundedness, one proves the well-known Cerulus-Martin lower bound on how fast an elastic scattering amplitude can decrease in the hard-scattering regime. 
In this paper we consider two non-trivial extensions of the previous bound. (i)~We generalize the assumption of polynomial boundedness by allowing  amplitudes to exponentially grow for some complex momenta and prove a more general lower bound in the hard-scattering regime.
(ii)~We prove a new lower bound on elastic scattering amplitudes in the Regge regime, 
in both cases of polynomial and exponential boundedness. 
A bound on the Regge trajectory for negative momentum transfer squared is also derived. We discuss the relevance of our results for understanding gravitational scattering at the non-perturbative level and for constraining ultraviolet completions. In particular, we use the new bounds as probes of non-locality in black-hole formation,  perturbative string theory, classicalization, Galileons, and infinite-derivative field theories, where both the polynomial boundedness and the Cerulus-Martin bound are violated.
\end{abstract}
\vspace{4.3cm}
\noindent\rule{5.5cm}{0.4pt}\\
$\,^\star$
\href{mailto:luca.buoninfante@su.se}{luca.buoninfante@su.se}\\	
$\,^\dagger$ \href{mailto:jtokuda@amethyst.kobe-u.ac.jp}{jtokuda@ibs.re.kr}\\
$\,^\ddagger$ \href{mailto:gucci@ibs.re.kr}{gucci@ibs.re.kr}

\end{titlepage}

{
	\hypersetup{linkcolor=black}
	\tableofcontents
}

\baselineskip=17.63pt



\newpage

\section{Introduction}

The S-matrix operator is a central object in particle physics, through which we can compute physical observables such as cross sections and decay rates. From its mathematical properties we can typically derive interesting physical implications. For example, the requirements of Lorentz invariance, polynomial boundedness, unitarity and analyticity can be sufficient conditions to guarantee the validity of several physical properties such as locality of interactions, conservation of probabilities, micro- and macro-causality~\cite{Weinberg:1995mt}. 

Given a Lagrangian in quantum field theory (QFT), we can compute the corresponding S-matrix and check whether the aforementioned mathematical and physical properties are verified. In the conventional framework of perturbative QFT without gravity those properties are indeed satisfied. However, some of the standard assumptions may be modified in the strong gravity regime or due to some non-perturbative effects (not necessarily of gravitational type). For instance, it is generally believed that the high-energy scattering in gravity could be dominated by black-hole formation which would cause an exponential suppression of the amplitudes and an exponential increase of the spectral density, thus implying a violation of polynomial boundedness~\cite{Banks:1999gd,Giddings:2007bw,Giddings:2007qq,Giddings:2009gj}. A similar violation of locality is expected to appear in some ultraviolet (UV) completions such as string theory at the perturbative level~\cite{Gross:1987kza,Gross:1987ar,Giddings:2006vu}, and because of non-perturbative effects in the classicalization proposal~\cite{Dvali:2010jz,Keltner:2015xda} and in Galileon theories~\cite{Keltner:2015xda}.

The assumption of specific fundamental principles for a given UV completion results in general properties of scattering amplitudes and, thus, imposes strong constraints on how the amplitudes behave in various kinematic regimes including the physical regions. These constraints are usually expressed in the form of upper or lower bounds~\cite{Yndurain:1972ix}. If such bounds are not respected, it means that at least one of the fundamental principles imposed at the level of  UV theory must be violated. Hence, these bounds can be very useful to gain information on the fundamental principles of  UV theory.

One of the most famous bounds in this context is the Froissart-Martin {\it upper} bound on a $2\rightarrow 2$ elastic scattering amplitude in the forward limit~\cite{Froissart:1961ux,Martin:1962rt}, that was derived under the assumption of polynomial boundedness in the unphysical domain in addition to unitarity and analyticity. 
Another important example is a {\it lower} bound on elastic scattering amplitudes in the high-energy limit and for fixed angle, that was derived in 1964 by Cerulus and Martin~\cite{Cerulus:1964cjb}. Their result relies on the assumptions of unitarity, analyticity, polynomial boundedness, and of a finite mass gap. In their formula, given an amplitude $\mathcal{M}(s,\cos \theta),$ $s$ being the center-of-mass energy squared and $\theta$ the scattering angle, the Cerulus-Martin bound reads~\cite{Cerulus:1964cjb}
\begin{equation}
\max_{-a\leq \cos\theta \leq a}\left|\mathcal{M}(s,\cos\theta)\right|\geq \mathcal{N}(s)\,e^{-f(a)\,\sqrt{s}\,\log (s/s_0)}\,,\label{C-M-bound}
\end{equation}
where $\mathcal{N}(s)$ is a positive function of $s$ that is subdominant in the $s\rightarrow\infty$ limit, $f(a)$ is a positive function of $a\in (0,1)$, and $s_0$ is some energy-squared reference scale. Note that, the above inequality is not a bound at some specific angle since any amplitude can in principle vanish at given angle. Instead, it represents a lower bound on the maximum of the modulus of the scattering amplitude for some finite interval of the scattering angle. Because only the scatttering at small impact parameters is relevant for the fixed-angle scattering, the Cerulus-Martin bound can be used as a clean probe of UV physics. 

In this work, we aim at investigating two non-trivial extensions of the Cerulus-Martin bound. (\textit{i})~Firstly, we replace the assumption of polynomial boundedness with the more general condition of \textit{exponential boundedness} that would be generically respected even in theories with non-locality. 
Thus, we derive a generalized lower bound on how fast an elastic scattering amplitude can decrease in the high-energy limit and for fixed angles (i.e., in the so-called \textit{hard-scattering limit}). 
We propose to parameterize the exponential boundedness and the degree of non-locality following  Jaffe's classification of strictly localizable, quasi-local and non-localizable theories~\cite{Jaffe:1966an,Jaffe:1967nb}. In this classification, the polynomial boundedness is included as a special case. 
(\textit{ii})~Secondly, we derive an analogue inequality in the high-energy limit and for fixed momentum transfer (i.e., in the so-called \textit{Regge limit}). This second derivation provides new rigorous lower bounds in both cases of polynomial and exponential boundedness. As a consequence, we also find a new bound on the Regge trajectory for negative values of momentum transfer squared.

Despite being quite general and interesting in their own,
our results can have important implications for 
constraining the degree of (non-)locality of the underlining theory in both gravitational and non-gravitational scenarios. We discuss applications to gravitational scattering processes at high energy, such as black-hole formation~\cite{Banks:1999gd,Giddings:2007qq,Arkani-Hamed:2007ryv,Dvali:2014ila}, and in the context of various UV completion scenarios. In particular, we use the new bounds as probes of non-locality by analysing the behavior of scattering amplitudes in perturbative string theory~\cite{Gross:1987kza,Gross:1987ar,Mende:1989wt}, classicalization proposal~\cite{Dvali:2010jz}, Galileon theories~\cite{Keltner:2015xda}, and infinite-derivative QFTs~\cite{Efimov:1967pjn,Krasnikov:1987yj,Tomboulis:1997gg,Buoninfante:2018mre}. In all these scenarios/theories the standard Cerulus-Martin bound~\eqref{C-M-bound} is violated. However, this does not mean a pathology because polynomial boundedness is not satisfied and, in fact, one has to verify whether the more general lower bound derived under the assumption of exponential boundedness is respected.

The present paper is organized as follows.
\begin{description}
	
\item[\textbf{sec.~\ref{sec:nonloc^2}:}] We introduce the idea of non-locality in field theory language and briefly review Jaffe's classification of strictly localizable, quasi-local and non-localizable QFTs. We explain our parametrization of (non-)locality in terms of the high-energy behavior of the scattering amplitude. 
In addition, we discuss examples of possible candidates for non-localizable theories. 

 \item[\textbf{sec.~\ref{sec:assump}:}] We briefly review basics of kinematics and introduce the working setup. We explain several mathematical details and introduce the main assumptions needed for the derivations of the lower bounds.
	
\item[\textbf{sec.~\ref{sec:lower-bound}:}] We extend the proof of Cerulus and Martin~\cite{Cerulus:1964cjb} by generalizing the assumption of polynomial boundedness with exponential boundedness. We obtain a more general lower bound on an elastic scattering amplitude in the hard-scattering limit.  

\item[\textbf{sec.~\ref{sec:lower-bound-Regge}:}] We prove a general lower bound on an elastic scattering amplitude in the Regge limit. 
As an implication we also obtain a bound on the Regge trajectory for negative values of the momentum transfer squared. 
	
\item[\textbf{sec.~\ref{sec:examples}:}] We use the newly derived bounds as probes of non-locality in gravity and in some UV completion scenarios. In particular,  
we consider several scenarios/theories in which the standard Cerulus-Martin bound can be violated, such as gravitational scattering and black-hole formation, classicalization proposal, Galileon theories, perturbative string theory, and infinite-derivative QFTs. 
	
\item[\textbf{sec.~\ref{sec:discuss}:}] This section is devoted to summary of the main results, conclusions and outlook.

\item[\textbf{sec.~\ref{app-C(x)}:}] We provide more mathematical details needed to understand some steps in the derivations of the lower bounds.
	
\end{description}

Throughout the whole paper we use natural units $\hbar=1=c$ and adopt the mostly positive convention for the metric signature $(\eta_{\mu\nu})={\rm diag}(-1,+1,+1,+1).$

\section{Locality, non-localizability, and S-matrix}\label{sec:nonloc^2}

In this work, we consider general S-matrices possessing 
certain features which are expected to capture the (non-)local nature of the underlying theory. We propose to quantify the degree of non-locality of a given theory by extending
the locality criterion introduced by Jaffe for the Wightman functions in~\cite{Jaffe:1966an,Jaffe:1967nb} to the behavior of the scattering amplitudes. We review Jaffe's criterion in sec.~\ref{subsec:nonlocalJ}. Then, we briefly explain our parameterization of non-locality in sec.~\ref{subsec:nonlocalS} (in  sec.~\ref{subsec:assumptions} we give a more precise parameterization). We 
introduce  several concrete scenarios in which non-local features are expected to be present in sec.~\ref{subsec:candidate}. 
To introduce Jaffe's locality criterion we are going to work with the simplest case of a massive scalar field. However, the same concept can also be generalized to massless fields and, indeed, in sec.~\ref{subsec:candidate} we will consider examples of gapless theories.

\subsection{Locality criterion by Jaffe}\label{subsec:nonlocalJ}

We now introduce the concepts of strictly localizable, quasi-local and non-localizable fields through Jaffe's classification~\cite{Jaffe:1966an,Jaffe:1967nb} according to which the local or non-local nature of a quantum field is incorporated in the properties of its $n$-point Wightman functions. 
We are going to present an intuitive understanding of Jaffe's idea of (non-)locality without providing a rigorous treatment of (non-)localizable QFTs which, in fact, is not required for the scope of this paper.

For illustration purposes, we focus on the two-point Wightman function of a scalar field $\phi$ with a physical mass $m$. We may formally write its expression in position space as 
\begin{align}
    W(x,y)
    =
    \bra{0}\phi(x)\phi(y)\ket{0}
    \,,
\end{align}
where $\ket 0$ denotes a Lorentz invariant vacuum, although strictly speaking this object is generally understood as a distribution. Let us consider an interacting QFT of $\phi$ which is subject to the Wightman axioms (such as Lorentz invariance, the spectral condition, etc) and whose Wightman functions are understood as tempered distributions~\cite{Wightman:1956zz}. Consequently, the two-point Wightman function of $\phi$ in momentum space is positive, Lorentz invariant, and supported in $V_+= \{p: -p^2\geq m^2, p^0>0\}$, satisfying 
\begin{align}
    \int\mathrm{d}^4p\,
    \frac{\tilde W(p)}{(1+\|p\|^2)^n} 
    < \infty
    \,,\label{eq:temp}
\end{align}
for some finite $n$. Here, $\tilde W(p)$  is the Wigthman function in momentum space and $\|p\|\coloneqq (\sum_{i=1}^4 p_i^2)^{1/2}$ denotes the Euclidean length of a real four-vector $p$. 
It may be useful to write $\tilde W(p)$ in terms of the K\"{a}ll\'{e}n-Lehmann spectral density $\rho(-p^2)$ as
\begin{align}
    \tilde W(p)
    = 2\pi\,\Theta(p^0)\rho(-p^2)\,,
\end{align}
where $\Theta(p^0)$ is the Heaviside step function and 
$\rho(-p^2)$ is required to be a non-negative distribution with support in $V_+$. In a free theory, we have $\rho(-p^2)=\delta(p^2+m^2)$. The condition~\eqref{eq:temp} requires that $\rho(-p^2)$ is polynomially bounded in the large momentum limit $-p^2\to\infty$.

Jaffe pointed out that the restriction on the growth rate of the momentum-space Wightman function~\eqref{eq:temp} can be relaxed without spoiling the notion of locality~\cite{Jaffe:1966an,Jaffe:1967nb}.\footnote{Technically, this generalization is done by restricting the space of test functions to the Gel'fand-Shilov space which is a subspace of the Schwartz space; the latter space is used in the original Wightman's formulation. Consequently, Wightman functions in position space are not necessarily tempered distributions in Jaffe's formulation, and the bound~\eqref{eq:temp} on the momentum-space Wightman function can be relaxed.} Let us define $\mu\equiv -p^2$ and parameterize the growth rate of $\rho(\mu)$ as follows
\begin{align}
    \rho(\mu)\leq A\,\mu^N e^{c\,\mu^\alpha}
    \,,\label{eq:growth}
\end{align}
where $N$ and $\alpha$ are real non-negative parameters, while $A$ and $c$ denotes some positive constants. According to the classification proposed by Jaffe, a field $\phi$ is said to be a strictly localizable field 
or a non-localizable field depending on the value of $\alpha$ as~\footnote{More precisely, 
the condition of strict localizability is given by~\cite{Jaffe:1966an,Jaffe:1967nb}
\begin{align}    \int^\infty_0\mathrm{dt}\,\frac{\ln\rho(t^2)}{1+t^2}
    <\infty
    \,.\label{eq:rigorousclass}
\end{align}
If we parameterize the growth rate of $\rho$ as \eqref{eq:growth} with $\mu=t^2$, the condition \eqref{eq:rigorousclass} gives $\alpha<1/2$.} 
\begin{equation}
\begin{cases} 
0\leq\alpha<\frac{1}{2}&:\,\,\text{strictly localizable}\,,\\[2mm]
\alpha\geq \frac{1}{2}&:\,\,\text{non-localizable}\,.
\end{cases}\label{eq:classific}
\end{equation}
Note that the $\alpha=1/2$ case is special and is also called quasi-local; the reason for this is explained below.
The $\alpha=0$ case is the conventional one in which the Wightman functions are understood as tempered distributions. The generalization of this classification to $n$-point Wightman functions $\tilde W(p_1,\dots, p_n)$ in momentum space is straightforward: 
\begin{align}
    |\tilde W(p_1,\dots,p_n)|
    \leq
    A
    \left(\sum_{i=1}^n\|p_i\|^2\right)^N
    \exp\left[c_n\left(\sum_{i=1}^n\|p_i\|^2\right)^{\alpha}\right]
    \,,\label{wight_growth1}
\end{align}
where $A$ and $c_n$ denote some other positive constants. The parameter $\alpha$ is defined such that these conditions are satisfied for all $n$ (for more details, see also~\cite{Iofa:1969fj,Iofa:1969ex}). Then, the degree of non-locality of the theory is defined according to~\eqref{eq:classific}.

To get some intuition of~\eqref{eq:classific}, following~\cite{Keltner:2015xda}, let us define the Wightman function $W(x,y)$ in position space by naively performing  the Fourier transform of $\tilde W(p)$ as 
\begin{align}
    W(x,y)
    =
    \int\frac{\mathrm{d}^4p}{(2\pi)^4}
    \tilde W(p) e^{ipz}
    =
    \int^\infty_0\mathrm{d}\mu\,\rho(\mu)
    W_\text{free}(z;\mu)
    \,,\label{eq:naivewight}
\end{align}
where $z\coloneqq x-y$ and $W_\text{free}(z;\mu)$ denotes a two-point correlation function in position space for a free field $\phi$ with mass $m=\sqrt{\mu}$, i.e.,
\begin{align}
    W_\text{free}(z;\mu)
    \coloneqq
    \int\frac{\mathrm{d}^4p}{(2\pi)^3}
    \Theta(p^0)\delta(p^2+\mu) e^{ipz}
    \,.\label{eq:freedwight}
\end{align}
From the asymptotic behavior of the free Wightman function at large $|z^2|$, i.e.,
\begin{eqnarray}
W_\text{free}(z;\mu)\sim \label{estimate_freewight}
\begin{cases}\displaystyle\frac{(2\sqrt{\mu})^{1/2}}{\bigl(4\pi \sqrt{z^2}\bigr)^{3/2}}e^{-\sqrt{\mu z^2}}& \text{for}\,\,\,\mu z^2\gg1\,,\\[3mm]
\displaystyle-ie^{-i\pi/4}\frac{(2\sqrt{\mu})^{1/2}}{\bigl(4\pi \sqrt{-z^2}\bigr)^{3/2}}e^{-i\sqrt{-\mu z^2}}&\text{for}\,\,\,-\mu z^2\gg1\,,
\end{cases}
\end{eqnarray}
we can notice that the integral in eq.~\eqref{eq:naivewight} does not converge even for $x\neq y$ when $\alpha\geq 1/2$ because the high-energy behavior of the spectral density dominates. This intuitively explains why $\alpha<1/2$ gives a locality criterion as in this case the Wightman function can be defined without the need of smearing the fields. The $\alpha=1/2$ case is special in the sense that~\eqref{eq:naivewight} is convergent for sufficiently large but finite $|x-y|$. This is the reason why $\alpha=1/2$ is called quasi-local; see also~\cite{Iofa:1969fj,Iofa:1969ex}.

In general, a field $\phi$ is understood as an operator-valued distribution which must be averaged with a smooth test function $f_{x_0}$ centered around some reference point $x_0,$ i.e., 
\begin{align}
    \phi[f_{x_0}]
    \coloneqq
    \int\mathrm{d}^4x\,
    \phi(x) f_{x_0}(x)
    \,,\label{eq:op}
\end{align}
whose Wightman function is everywhere finite including the coincident point $z=0$. Here, $f_{x_0}$ is defined in terms of a smooth function $\tilde f(p)$ as
\begin{align}
    f_{x_0}(x)
    \coloneqq
    \int\frac{\mathrm{d}^4p}{(2\pi)^4}
    \,\tilde f(p) e^{ip(x-x_0)}
    \,.\label{eq:testfn1}
\end{align}
In what follows, we assume that $f(x)$ is real for any real four vector $x,$ thus we have $\tilde f(-p)=\tilde f^*(p)$.
Now we define a two-point Wightman function of an operator $\phi[f_x]$ as 
\begin{align}
    W[f_x,f_y]
    \coloneqq
    \bra{0}\phi[f_x]\phi[f_y]
    \ket{0}
    &=
    \int\frac{\mathrm{d}^4p}{(2\pi)^3} \, 
    \Theta(p^0) |\tilde f(p)|^2 \rho(-p^2)
    e^{ip(x-y)}
    \,.\label{Wdef}
\end{align}
This expression is ensured to be finite when $|\tilde f(p)|<e^{-c\| p\|^{2\alpha}/2 } \times \text{(subdominant pieces)}$ in the large momentum limit $\| p\|\to\infty$. Note that $-p^2\to\infty$ implies $\| p\|\to\infty$.

A test function $f(x)$ has a compact support if and only if $\alpha<1/2$~\cite{Jaffe:1967nb}.
To understand this, let us define $g(x)$ as the Fourier transform of a smooth function $\tilde g(p)$,
\begin{align}
    g(x)
    =\int^\infty_{-\infty}\frac{\mathrm{d}p}{2\pi}\,
    \tilde g(p)e^{ipx}
    \,,
    \label{eq:gdef}
\end{align}
and assume that $\tilde g(p)$ is bounded on the real axis as $|\tilde g(p)|<b\, e^{-c| p|^{2\alpha}}$ for $p>L$ and $p<-L$ with a sufficiently large but finite $L>0$. Here, $b$ and $c$ denote positive constants. When $\alpha>1/2$, $g(x)$ is also analytic in the complex $x$-plane because it can be shown to be finite for any $x\in\mathbb C$:
\begin{align}
    |g(x)|
    <
    \left|
    \int^L_{-L}\frac{\mathrm{d}p}{2\pi}\,
    \tilde g(p)e^{ipx}
    \right|
    +
    b\int^\infty_{L}\frac{\mathrm{d}p}{\pi}\,
    e^{-c|p|^{2\alpha}}e^{-p\im(x)}
    +
    b\int^{-L}_{-\infty}\frac{\mathrm{d}p}{\pi}\,
    e^{-c|p|^{2\alpha}}e^{-p\im(x)}
    < \infty
    \,.
\end{align}
Similarly, it is also possible to show that $g(x)$ is analytic in the strip $\{x: -\infty<\re(x)<\infty, -c<\im(x)<c\}$ when $\alpha=1/2$. 
In the $\alpha\geq 1/2$ case, the analyticity domain of $g(x)$ includes the whole real axis. Hence, we conclude that any nontrivial test function $g(x)$ on the real axis cannot be compactly supported when $\alpha\geq 1/2$.
A simple example of a function with non-compact support is the Gaussian  $\tilde g(p)\sim e^{-c\,p^2}$ which corresponds to the $\alpha=1$ case. Its Fourier transform is also a Gaussian $g(x)\sim e^{-x^2/4c}$ which is entire and not compactly supported.

This explanation indicates that the Fourier transform of a function with compact support on the real axis cannot decrease arbitrarily fast. 
In the above discussion, it was shown that $|\tilde g(p)|$ cannot decrease faster than $e^{-c|p|}$ along the real axis if its Fourier transform $g(x)$ is compactly supported.  Interestingly, this behavior for large $|p|$ is similar to what happens with the Cerulus-Martin lower bound~\eqref{C-M-bound} on the $2\rightarrow 2$ scattering amplitude in the hard-scattering regime, i.e.,
\begin{align}
    |\scat(s,\cos\theta)|
    \gtrsim e^{-c\sqrt{s}}
    \,,\label{eq:testlocal}
\end{align}
which was derived for local ($\alpha=0$) QFTs~\cite{Cerulus:1964cjb}.\footnote{However, according to our proof in sec.~\ref{sec:lower-bound}, when $\alpha>0$ the lower bound gets smaller by an additional factor $s^\alpha$ in the exponent, even in the case of strictly localizable theories (see eq.~\eqref{new-bound}). It would be interesting to understand this discrepancy between the intuition based on the property of compactly-supported test functions and the bound \eqref{new-bound}; for instance, the intuition may suggest the existence of a bound which is stronger than \eqref{new-bound} in the strictly localizable case. We leave this point for a future work.}

\paragraph{S-matrix.}
It is worth mentioning that the formulation of quasi-local and non-localizable theories has been developed. The Wightman formulation of these QFTs was given  
in Refs.~\cite{Iofa:1969fj,Iofa:1969ex}. Also, under some working assumptions, in Ref.~\cite{Steinmann:1970cm} it was proposed that one can construct a Lorentz invariant and unitary S-matrix which exhibits standard properties such as cluster decomposition, LSZ formalism, and CPT invariance. 

\paragraph{Commutation rules.} 
Causality of local QFT with $\alpha=0$ is usually ensured by the local commutation rule
\begin{align}
    [\phi(x),\phi(y)]
    = 0
    \quad \text{for} 
    \quad
    (x-y)^2>0
    \,.\label{localcommute}
\end{align}
This can also be expressed in terms of the Wightman function in position space. For the two-point function, eq.~\eqref{localcommute} implies 
\begin{align}
    W(x-y)
    - W(y-x)
    = 0
    \quad \text{for} 
    \quad
    (x-y)^2>0
    \,.
    \label{commute_wightman}
\end{align}
where $W(x-y)\equiv W(x,y)$.
The generalization to $n$-point functions is straightforward; see~\cite{Wightman:1956zz} for more details. For strictly localizable QFTs with $0<\alpha<1/2$, it will be straightforward to impose the local commutativity condition~\eqref{localcommute} because the test functions can have compact support~\cite{Jaffe:1966an,Jaffe:1967nb}.

For non-localizable QFTs with $\alpha\geq1/2$, instead, we expect that the commutator $[\phi(x),\phi(y)]$ does not vanish even for space-like separated spacetime points due to the non-local nature of the theory, i.e.,
\begin{equation}
\left[\phi(x),\phi(y)\right]\neq 0\quad {\rm for}\,\,\,(x-y)^2>0\,.\label{viol-micro}
\end{equation}
In such a case, it will be necessary to modify~\eqref{commute_wightman}.
Instead of~\eqref{commute_wightman}, we need to consider the following commutator smeared by some test function $f$ with non-compact support:
\begin{align}
    W[f_x,f_y]-W[f_y,f_x]
    =
    \int\mathrm{d}^4\xi\,
    \left[
        W(\xi)-W(-\xi)
    \right]
    T^{f}_{x-y}(\xi)
    \,,\label{commute_nonlocal}
\end{align}
where $T^{f}_{x-y}(\xi)$ denotes a test function defined in terms of an original test function $f$ as 
\begin{align}
    T^{f}_{x-y}(\xi)
    \coloneqq
    \int\frac{\mathrm{d}^4p}{(2\pi)^4}\,
    \left|\tilde f(p)\right|^2
    e^{ip\left(\xi-(x-y)\right)}
    \,.
\end{align}
Because $T^{f}_{x-y}(\xi)$ will be nonzero for some timelike $\xi$ even 
when $(x-y)^2>0$ in general, we expect that the smeared commutator \eqref{commute_nonlocal} does not vanish even for spacelike separated points $(x,y)$. Nevertheless, one can expect that \eqref{commute_nonlocal} decays exponentially in the limit $(x-y)^2\to\infty$ because $|T^{f}_{x-y}(\xi)|$ decays exponentially for large $\|\xi-(x-y)\|$. For instance, when $\alpha=1$ we have $|\tilde f(p)|^2\sim e^{-\sigma\| p\|^{2} }$, thus $|T^{f}_{x-y}(\xi)|$ decays as fast as a Gaussian $e^{-\sigma \|\xi-(x-y)\|^2}$.

Indeed, for the nonlocalizable case with $\alpha>1/2$, Ref.~\cite{Fainberg:1992jt} proposed an \textit{asymptotic} commutativity condition instead of the local commutation rule~\eqref{localcommute} or \eqref{commute_wightman}. Roughly speaking, the condition reads
\begin{align}
    \left|
    \int \mathrm{d}^4\xi \,
    \left[W(\xi)-W(-\xi)\right]
    T^f_{x-y}(\xi)
    \right|
    \lesssim 
    A\, e^{-\sigma\|x-y\|^{2\alpha/(2\alpha-1)}}
    \quad \text{for} 
    \quad 
    (x-y)^2>0
    \,,\label{asymptotic_commute}
\end{align}
where $A$ and $\sigma$ denote some positive constants whose precise values are determined by the functional form of $T^f_{x-y}(\xi)$.
The condition \eqref{asymptotic_commute} is imposed for all the allowed test functions $T^f_{x-y}(\xi)$ for a given non-localizable theory. For a more precise definition of asymptotic commutativity and its implications see~\cite{Fainberg:1992jt,Soloviev:1999rv} and references therein.

In the case $\alpha=1/2$, the condition \eqref{asymptotic_commute} can be improved thanks to the quasi-local nature of the theory: $T^f_{x-y}(\xi)$ can vanish for some $\xi$ in this case. In particular, when $T^f_{x-y}(\xi)=0$ for any $\xi$ satisfying $\xi^2
\leq\ell^2$, the quasi-locality (or ``$\ell$-locality") condition holds~\cite{Iofa:1969fj},
\begin{align}
    \int \mathrm{d}^4\xi 
    \left[
        W(\xi)
        - W(-\xi)
    \right]
    T^f_{x-y}(\xi)
    = 0 
    \,.
    \label{ell_locality}
\end{align}
Here, $\ell$ is defined by the growth rate of spectral density as $\ell/\sqrt{2} = {\displaystyle\lim_{t\to\infty}}\ln\rho(t^2)/t$. 
Since $T^f_{x-y}(\xi)$ cannot vanish at $\xi=x-y$, the quasi-local commutativity \eqref{ell_locality} can be satisfied only when $x$ and $y$ are sufficiently separated to satisfy $(x-y)^2>\ell^2$. This result is natural from the perspective that the non-locality appears only up to a certain finite distance $\ell$ for quasi-local case as
already mentioned below eq.~\eqref{estimate_freewight}. The generalization of this condition to $n$-point functions is also given in~\cite{Iofa:1969fj}.

\paragraph{Connection with low-energy EFTs.}
Even though the smeared commutator \eqref{commute_nonlocal} does not vanish even when $x$ and $y$ are space-like separated in non-localizable theories ($\alpha\geq1/2$), it becomes negligibly small for $(x-y)^2\gg \sigma^{(1/\alpha)-2}$. In this sense, the parameter $\sigma$ 
may govern the scale of non-locality. This suggests that, even when a UV completion is non-localizable, its effective description at low energies below a certain non-locality scale would be given in terms of a local QFT for which the usual local commutation rule~\eqref{localcommute} is satisfied. This point may also be clarified by performing the low-energy expansion of the exponentially-growing spectral density $\rho(\mu)$. For instance, the function $\rho(\mu)=\mu^Ne^{\mu/\mu_0}$ can be well approximated by polynomials of some finite degree when focusing on the sufficiently low-energy regime $\mu\ll\mu_0$. This means that, as far as low-energy regimes are concerned, the theory can be well approximated by the standard local QFT with $\alpha=0;$ see~\cite{Keltner:2015xda} for more details on the EFT perspective of non-localizability.

\paragraph{Green's functions.} 
For usual local QFTs with $\alpha=0$, the time-ordered two-point function
admits the K\"{a}ll\'{e}n-Lehmann spectral representation which is analogous to \eqref{eq:naivewight} for the Wightman function. Let us suppose that $\rho(\mu)$ decays faster than $1/\log\mu$  for $\mu\to\infty$ as a toy example. In this case, the Wightman function in position space is everywhere finite and \eqref{eq:naivewight} is correct. Similarly, the spectral representation for the  time-ordered two-point function $\tilde D_F(p)$  in momentum space is given by~\cite{Steinmann:1963}
\begin{align}
    \tilde D_F(p)
    =
    \int^\infty_0\mathrm{d}\mu\,
    \rho(\mu)
    \frac{-i}{p^2+\mu-i\epsilon}
    \,.\label{eq:feynspectral0}
\end{align}
Substituting $\rho=\delta(\mu-m^2)$ into this equation, we recover $\tilde D_F(p)$ in the free theory. When $\rho$ grows polynomially as $\rho(\mu)\sim \mu^N$, eq.~\eqref{eq:feynspectral0} diverges. In such cases, the spectral representation is derived as~\cite{Steinmann:1963}
\begin{align}
    \tilde D_F(p)
    =
    Q_{N+1}(-p^2)\int^\infty_0\mathrm{d}\mu\,
    \frac{\rho(\mu)}{Q_{N+1}(\mu)}
    \frac{-i}{p^2+\mu-i\epsilon}
    + i\sum_{j=0}^N c_j (p^2)^j
    \,.\label{eq:feynspectral1}
\end{align} 
Here, a positive polynomial $Q_{N+1}(-p^2)$ of $(N+1)$-th degree is introduced such that the integral converges; $c_j$ are some constants. The choice of $Q_{N+1}$ is not unique. This freedom leads to an ambiguity in the definition of $\tilde D_F$ which can be renormalized by $c_j$. In position space, these ambiguous terms correspond to the sum of derivatives of the delta function. 
We can similarly define the spectral representation for $\tilde D_F(p)$ even in the non-standard case $\alpha>0$ where $\rho(\mu)\sim \mu^Ne^{c\mu^\alpha}$ at large $\mu$. In such cases, we should use a positive function $Q(-p^2)$ which grows as fast as $(-p^2)^{N+1}e^{c(-p^2)^\alpha}$ instead of a polynomial $Q_{N+1}(-p^2)$.

The spectral representation~\eqref{eq:feynspectral1} suggests that when the Wightman functions grow in the large momentum limit, the time-ordered Green's function would also grow accordingly. For instance, when the former grows polynomially as $\rho(\mu)\sim\mu^N$ at large $\mu$, the growth rate of the latter may be bounded in the {\it complex} $p^2$-plane as 
\begin{align}
    \lim_{|p^2|\to\infty}
    \frac{\tilde D_F(p)}{|p^2|^{N+1}}
    =
    0\,.
\end{align}
A similar equation can also be written for $\alpha>0$ in which case the time-ordered Green's function may be  exponentially bounded in the complex $p^2$-plane.
 
The above discussion can also be extended to $n$-point functions with $n>2$. In particular, Ref.~\cite{Tokuda:2019nqb} derives the spectral representation for the time-ordered four-point Green's function $\tilde D_F^{(4)}(p_1,p_2,p_3,p_4)$ in momentum space, which suggests that the boundedness property of $\tilde D_F^{(4)}$ in the complex momentum plane will be determined by the growth rate of the four-point Wightman function $\tilde W(p_1,\cdots,p_4)$. When $\tilde W(p_1,\cdots,p_4)$ grows polynomially so that \eqref{wight_growth1} is satisfied with $\alpha=0$, for instance, the spectral representation for $\tilde D_F^{(4)}$ implies
\begin{align}
    \lim_{|-(p_1+p_2)^2|\to\infty}
    \left|
    \frac{\tilde D_F^{(4)}(-(p_1+p_2)^2,-(p_1-p_3)^2,\{p_i^2\})}
    {|-(p_1+p_2)^2|^{N+1}}
    \right|
    =0
    \,.\label{four_Feyn_bound}
\end{align}
Here, we used the fact that $\tilde D_F^{(4)}$ can be regarded as a function of $-(p_1+p_2)^2$, $-(p_1-p_3)^2$, and $\{p_i^2\}_{i=1,2,3,4}$. 
A similar equation can be also written for $\alpha>0$ in which case the Green's function would be exponentially bounded.

Because $\tilde D_F^{(4)}$ is related to the four-point scattering amplitude $\scat$ via the reduction formula,~\eqref{four_Feyn_bound} 
indicates that the four-point scattering amplitudes, which are related to the four-point time-ordered Green's function via the reduction formula, are also bounded in the large momentum limit accordingly. 
Indeed, it is known that the four-point scattering amplitude is polynomially bounded in the case $\alpha=0$, and this feature leads to the dispersion relations with a finite number of subtractions.  It is also found that, even for the $\alpha>1/2$ case, the upper bound on the four-point amplitude is given by the growth rate of the Wightman function as $|\scat(s,t)|<e^{s^{\alpha}}$ in the Regge limit ($s\to\infty$ with fixed $t$)~\cite{Fainberg:1971ia}. 
Here, $s$ and $t$ are usual Mandelstam variables corresponding to the center-of-mass energy squared and the momentum transfer squared, respectively.

\subsection{Locality criterion by S-matrix}\label{subsec:nonlocalS}

Based on the above observations, in this paper we {\it propose} to parameterize the non-locality of given models in terms of the growth rate of the scattering amplitude in the complex Mandelstam variable plane. We discuss the implications of such a rapid growth in the next sections. In particular, roughly speaking, our parameterization is the following:
\begin{align}
    |\scat(s,\cos\theta)|
    \leq
   A\, s^N\,e^{c \, s^\alpha}
    \quad
    \text{for}
    \quad s\to\infty
    \,,\label{eq:basicassump}
\end{align}
where $\cos\theta$ takes values in a certain domain of the complex $\cos\theta$-plane that we will define in the next section; e.g., see Fig.~\ref{fig1}. In  sec.~\ref{subsec:assumptions} we will provide more details on our assumptions that are needed to prove the lower bounds in secs.~\ref{sec:lower-bound} and~\ref{sec:lower-bound-Regge}.

Even though our parameterization of non-locality in eq.~\eqref{eq:basicassump}  is motivated by the Jaffe's criterion of (non-)localizability, we emphasize that our formulation does not explicitly rely on  Jaffe's classification of QFTs as our analysis is solely based on the properties of the S-matrix.
This means that, in principle, our study could be relevant even for theories which do not fall into Jaffe's QFTs but whose S-matrix has some peculiar features such as the non-polynomial behavior in eq.~\eqref{eq:basicassump}. 

\paragraph{Relation to causality.}
The macro-causality condition \eqref{ell_locality} or \eqref{asymptotic_commute} for non-localizable theories implies that the retarded Green's function $G_\text{ret}(x,y),$ which is defined under the appropriate smearing by test functions, can be non-zero for space-like separations but it must be sufficiently suppressed when $(x-y)^2$ is sufficiently larger than the square of the non-locality scale. Interestingly, by considering the $(1+0)$-dimensional scattering model (the so-called signal model discussed previously, 
 e.g., in~\cite{Eden:1971fm,Camanho:2014apa}), Ref.~\cite{Keltner:2015xda} claims that the analyticity of the S-matrix in the complex energy plane and the exponential boundedness~\eqref{eq:basicassump} can be sufficient to ensure the macro-causality for non-localizable theories; whereas analyticity and the  boundedness~\eqref{eq:basicassump} with $0\leq\alpha<1/2$
can imply causality for local and strictly localizable theories, i.e., eq.~\eqref{localcommute}. 
These observations motivate us to assume  suitable analytic properties and the condition of exponential boundedness~\eqref{eq:basicassump} on the behavior of the S-matrix.

\subsubsection{Positivity bounds on the S-matrix}\label{subsubsec:positivity}
The high-energy behavior of the scattering amplitudes in non-localizable theories was first discussed in~\cite{Fainberg:1971ia}. More recently, in Ref.~\cite{Tokuda:2019nqb} it was shown that under some working assumptions the  behavior of a scattering amplitude  $\mathcal{M}(s,t\leq0)$ in the Regge limit 
is polynomially bounded as
\begin{align}
    \left|\mathcal{M}(s,t\leq0)\right|< s\left(A\,s^{2\alpha}+B\,s^\alpha \log(s/s_0)+C\,\log^2(s/s_0)\right)\,,\label{eq:pbbound}
\end{align}
where $A,$ $B,$ $C$ are some positive constants and $s_0$ some energy scale squared; the Froissart bound~\cite{Froissart:1961ux} is consistently recovered in the local $\alpha=0$ case.
Combined with the Phragm\'{e}n-Lindel\"{o}f theorem, the bound~\eqref{eq:pbbound} gives rise to the fixed-$t$ dispersion relations for the $\alpha<1$ case with a finite number of subtractions, under the assumption of the sub-exponential growth of $|\scat(s,t)|$ in the limit $|s|\to\infty$ in the cut $s$-plane. In particular, the number of subtractions can be two in the strictly localizable case $\alpha<1/2$.\footnote{This does not mean that the twice-subtracted dispersion relation has to be always violated in the $\alpha\geq1/2$ case.} The twice-subtracted dispersion relation leads to the so-called positivity bound on the low-energy coefficient $c_{2,0}$ as~\cite{Pham:1985cr,Adams:2006sv}
\begin{align}
    c_{2,0}
    =
    \frac{2}{\pi}\int^\infty_{4m^2}\mathrm{d}s\,
    \frac{\im\,\scat(s,0)}{s^3} 
    >0
    \,,\label{eq:LOpositivity}
\end{align}
where $\{c_{n,m}\}$ are defined by 
\begin{align}
    \scat(s,t)
    =
    \text{($s$,$t$,$u$)-poles}
    +
    \sum_{n,m=0}^\infty c_{n,m} s^n t^m
    \,;
\end{align}
the condition $\im\,\scat(s,0)>0$ implied by unitarity is used in \eqref{eq:LOpositivity}.
Hence, EFTs with negative $c_{2,0}$ cannot be embedded into strictly localizable UV completions~\cite{Tokuda:2019nqb}. Recently, stronger bounds on low-energy coefficients have been derived by making use of crossing symmetry~\cite{Arkani-Hamed:2020blm,Bellazzini:2020cot,Caron-Huot:2020cmc,Tolley:2020gtv,Sinha:2020win}. As suggestive observations, we will see in sec.~\ref{subsec:candidate} that the positivity bound on $c_{2,0}$ derived from the twice-subtracted dispersion relation is indeed violated in some candidates for non-localizable theories (with $\alpha\geq 1/2$). 

On the other hand, the bound \eqref{eq:pbbound} also shows that it is possible to derive dispersion relations with a finite number of subtractions even for $1/2\leq\alpha<1$. This means that it is still possible to derive positivity bounds on higher-order coefficients such as $c_{n,0}$ with $n\geq4$ for theories with $1/2\leq \alpha <1$~\cite{Tokuda:2019nqb}. It would be also interesting to constrain non-localizable theories by using other methods such as the S-matrix bootstrap~\cite{Paulos:2016fap,Paulos:2016but,Paulos:2017fhb}. We leave this aspect for a future work.

\subsection{Candidates for non-localizable theories}\label{subsec:candidate}

Let us now discuss some examples of possible candidates for  non-localizable QFTs; see also~\cite{Keltner:2015xda} for a suggestive summary of these examples.  

\subsubsection{Gravity and black-hole formation}\label{sec:cand-gravity}

Black-hole (BH) physics provides a natural explanation to understand why gravity is inherently non-local. Let us focus on the $2\rightarrow 2$ scattering amplitude $\scat(s,t)$ of gravitons.
In the usual perturbative QFTs, it is possible to probe 
arbitrary short-distance scales $L\sim E^{-1}$ by considering high-energy hard scattering processes with $s,-t\sim E^2$. In Einstein's General Relativity (GR), however, there exists a lower limit on the distance scale $L$ that can be probed before BH formation sets in, and this is given by $L\gtrsim r_s(E)=2E/M_p^2\,.$ Here, $r_s(E)$ is the Schwarzschild radius of a BH of mass $E$. BH formation prevents us
from probing distances smaller than $r_s(E)$. The important feature is that the length scale $r_s(E)$ grows with the energy, thus more energetic probes will be affected by a larger uncertainty in resolving distances. 

In this case, the non-local nature of the theory may also be understood from the exponential growth of the spectral density. In fact, if BH states dominate at high energy, then the high-energy behavior of the spectral density will be proportional to the number of BH states~\cite{Banks:1999gd,Giddings:2007qq}, i.e. 
\begin{equation}
\rho(s)\sim e^{S_{\rm BH}(\sqrt{s})}=e^{c\,(\sqrt{s}/M_p)^{\frac{D-2}{D-3}}}\,,
\end{equation}
where $S_{\rm BH}\sim (\sqrt{s}/M_p)^{\frac{D-2}{D-3}}$ is the Bekenstein-Hawking entropy and $D$ is the spacetime dimension. Comparing this with Jaffe's criterion given by eqs.~\eqref{eq:growth} and \eqref{eq:classific}, we may identify the degree of non-locality as
\begin{equation}
\alpha=\frac{1}{2}\frac{D-2}{D-3}\,,\label{BHalpha}
\end{equation}
which depends on the spacetime dimensions $D$.
Interestingly, we have $\alpha>1/2$ for $D>3$. This implies that the gravitational theory is non-localizable. In particular, in $D=4$ we get $\alpha=1.$ 

The violation of locality can also be understood in the language of S-matrix around flat spacetime. In fact, it is believed that a $2\rightarrow 2$ scattering amplitude at high energies will be dominated by BH production which would make the scattering amplitude exponentially suppressed. For instance, for fixed-angle (hard-scattering regime) one expects~\cite{Arkani-Hamed:2007ryv,Dvali:2014ila}:\footnote{It is worth mentioning that the behavior in eq.~\eqref{suppressed-BH} may not only be typical of Einstein's GR, but it might also appear in some renormalizable higher-derivative theories of gravity. The reason for this is that the tree-level amplitudes involving massless gravitons as external legs are the same as in GR~\cite{Dona:2015tra,Abe:2022spe,Holdom:2021hlo}; however, see~\cite{Holdom:2021hlo,Holdom:2022npq} for a different point of view.
On the other hand, in some other approaches to quantum gravity, graviton-graviton  amplitudes could still be polynomially bounded and respect the Cerulus-Martin bound; e.g., this was claimed in the context of asymptotically safe gravity~\cite{Draper:2020bop}. It would be interesting to understand what happens to BH formation in this case; e.g., see~\cite{Bosma:2019aiu,Platania:2023srt} for some discussions.}
\begin{equation}
E\gg M_p\quad \Rightarrow \quad \mathcal{M}(s,\cos\theta)\sim e^{-S_{\rm BH}(\sqrt{s})}=e^{-c\,(\sqrt{s}/M_p)^{\frac{D-2}{D-3}}}\,,
\label{suppressed-BH}
\end{equation}
where $\sqrt{s}$ is the center-of-mass energy of the incoming two-particle state. In sec.~\ref{sec:examp-BH-GR} we will explicitly prove that the behavior~\eqref{suppressed-BH} implies that $\scat(s,t)$ violates polynomial boundedness.

A few remarks are in order. First, the above estimates of the contributions from BH formation to $\scat(s,t)$ are suggestive, but to our knowledge such non-perturbative contributions have not been rigorously evaluated yet. There is a possibility
that gravitational amplitudes $\scat(s,t<0)$ with fixed $t$ are actually polynomially bounded in any directions in the upper half complex $s$-plane.  Second, even though the above estimates indicate some non-polynomial behavior of the amplitude, we can expect that $\scat(s,t<0)$ with fixed $t$ is polynomially bounded in any directions in the upper half-plane in higher spacetime dimensions $D>4$. The reason for this is the following. In $D>4$, partial-wave unitarity would imply the polynomial boundedness of $\scat(s,t<0)$ along the real-$s$ axis under some physical assumptions~\cite{Haring:2022cyf}, though we may have the exponential growth $\scat(s,t<0)\sim e^{|s|^{\alpha}}$ with eq.~\eqref{BHalpha} in some directions in the upper half-plane. Because we expect $\alpha<1$ when $D>4$, the Phragm\'{e}n-Lindel\"{o}f theorem 
implies that $\scat(s,t<0)$ is polynomially bounded in any directions in the upper half-plane. The discussion here is consistent with Ref.~\cite{Haring:2022cyf} in which the high-energy behavior of gravitational amplitudes is investigated. 

Finally, in the S-matrix language the BH formation is described as a non-perturbative scattering process and
can be effective 
only in the super-Planckian regime $s\gg M_p^2.$ 
If gravity is UV completed within its weakly-coupled regime $s\lesssim M_p^2$, it is not necessary to concern about BH formation and we may be interested in UV complete amplitudes in which higher-order terms in $1/M_p^2$ are truncated. We can expect that such amplitudes would satisfy the fixed-$t$ dispersion relation with finite number of subtractions: e.g., tree-level string amplitudes satisfy them.

\subsubsection{Classicalization proposal}\label{sec:cand-classicaliz}

In Ref.~\cite{Dvali:2010bf}, the softening behavior of scattering amplitudes caused by BH formation was proposed as a mechanism to achieve the UV completion of GR. The same type of idea was generalized to other non-renormalizable theories in which the high-energy behavior of scattering amplitudes gets weakened by the formation of a macroscopic semi-classical object called classicalon (analog of the BH). This type of UV completion was named \textit{classicalization}~\cite{Dvali:2010jz}.

An example of classicalizing theory was claimed to be the Goldstone model with massless kinetic term and derivative potential~\cite{Dvali:2012zc}. Let us consider the following scalar field theory 
\begin{align}
    \mathcal{L}
    =
    -\frac{1}{2}(\der\phi)^2 
    + \frac{\epsilon}{\Lambda^4}(\der\phi)^4
    +\cdots
    \,,\label{eq:model1}
\end{align}
where $\epsilon=\pm1$ and the ellipses stands for the higher-order terms. This model violates perturbative unitarity at $s\sim\Lambda^2$, suggesting the appearance of new degrees of freedom at around this scale. According to the standard Wilsonian UV completion point of view, such degrees of freedom are some new heavy particles. The model~\eqref{eq:model1} is then regarded as an EFT.
A famous example of this type of scenario is the UV completion of electroweak theory by the Higgs boson, where locality is mantained also in the UV. Hence, the positivity bound~\eqref{eq:LOpositivity} which requires $\epsilon=1$ for the model~\eqref{eq:model1} should be satisfied and, in known examples, this is indeed the case; see also~\cite{Adams:2006sv} for other examples.

However, in Ref.~\cite{Dvali:2010jz} another possibility for UV completion was proposed, according to which   
the scalar field $\phi$ remains the only perturbative degree of freedom at high energies, but new semi-classical states can be formed non-perturbatively as a collective phenomena of $\phi$-particles.
In this non-Wilsonian scenario, the scattering amplitude is unitarized through the formation of a classicalon which is a macroscopic bound state of $\phi$-particles. This is analogous to the BH formation in GR and its radius grows as the energy $\sqrt{s}$ increases with a certain positive power that is model-dependent. As a result, distances shorter than the size of the classicalon cannot be probed, thus classicalization is characterized by some inherent non-locality. Then, we expect that the S-matrix of the classicalizing models has some peculiar features reflecting non-locality. 

In fact, because of classicalon formation, a $2\rightarrow 2$ scattering amplitude in the model~\eqref{eq:model1} is expected to be exponentially suppressed in the hard scattering regime as $\scat(s,\cos\theta)\sim e^{-c\,s^{2/3}}$. In addition, the entropy of the classicalon was claimed to be given by $S\sim c\,s^{2/3}$~\cite{Dvali:2010jz}. 
As a consequence, the high-energy behavior of the spectral density is expected to be dominated by the classicalon density of states, i.e.,
\begin{equation}
\rho(\mu)\sim e^{c\,\mu^{2/3}}\,.
\end{equation}
This indicates that the classicalizing Goldstone model in eq.~\eqref{eq:model1} will be a non-localizable QFT with $\alpha=2/3.$ Interestingly, it was found in Ref.~\cite{Dvali:2012zc} that the classicalon solution exists only when $\epsilon=-1$, which violates the positivity bound~\eqref{eq:LOpositivity}. This is consistent with the finding in Ref.~\cite{Tokuda:2019nqb} according to which the violation of~\eqref{eq:LOpositivity} excludes the strictly localizable QFTs ($\alpha<1/2$) as candidates for UV completion. 

More generally, the classicalon solution depends on the details of the higher-derivative interaction terms. We refer to the $n$-point vertex with $2k$ derivatives as $\partial^{2k}\phi^n$; e.g., the quartic derivative term $(\der\phi)^4$ corresponds to $(k,n)=(2,4)$. Given an interaction term $\partial^{2k}\phi^n,$ it was argued that the density of states of the classicalon should be proportional to $e^{c\,r_*\sqrt{s}},$ where $r_*\sim (\sqrt{s})^{\frac{n-2}{n+4k-6}}$ is the size of the classicalon (also called classicalization radius) in $D=4$ spacetime dimensions~\cite{Dvali:2010jz}. Notice that the largest radius corresponds to the lowest order interaction term in a derivative expansion in the Lagrangian. Thus, in a classicalizing model whose lowest order interaction term contains $2k$ derivatives and $n$ fields, we expect the spectral density to grow as
\begin{equation}
\rho(\mu)\sim e^{c\,\mu^\alpha}\,,\quad 
\alpha=\frac{n+2k-4}{n+4k-6}\,.
\label{alpha-classicaliz}
\end{equation}
From the expression of $\alpha,$ it follows that $\alpha>1/2$ for $n\geq 3,$ which means that classicalizing theories are always non-localizable. Note that, in the case $k=1$ we have $\alpha=1$ for any $n,$ in agreement with four-dimensional Einstein's gravity.

\subsubsection{Galileon theories}\label{sec:cand-galil}
Ref.~\cite{Keltner:2015xda} argued that the Galileon model belongs to the class of non-localizable QFTs with $\alpha=3/5$ in $D=4$ spacetime dimensions. In particular, the spectral density is found to grow exponentially as $\rho(\mu)\sim e^{c\,\mu^{3/5}}$ in the Galileon theory which can be mapped to the free theory via Galileon duality~\cite{deRham:2013hsa}. 
This behavior may be also understood consistently from the viewpoint of the classicalization proposal as suggested in~\cite{Keltner:2015xda}. The Lagrangian of Galileon theories is~\cite{Nicolis:2008in}
\begin{equation}
    \mathcal{L}
    =
    -\frac{1}{2}(\partial\phi)^2
    +\frac{g_3}{\Lambda^3}(\Box\phi)(\partial\phi)^2
    +\frac{g_4}{\Lambda^4}(\der\phi)^2
    \left[
        (\Box\phi)^2-(\der_\mu\der_\nu\phi)^2
    \right]
    +\cdots\,,
\label{cubic-galileon}
\end{equation}
where $g_3$ and $g_4$ are dimensionless coupling constants, and the higher-order terms are suppressed. The second term 
and the third term are the lowest-order interaction terms called the cubic Galileon and the quartic Galileon, respectively. These non-renormalizable interactions lead to the violation of perturbative unitarity at energy scales above $\Lambda$.  Since the $(\der\phi)^4$ term is absent in the Lagrangian, the positivity bound~\eqref{eq:LOpositivity} derived from the twice-subtracted dispersion relation is violated. Even though the positivity bound is applicable only for gapped theories, strictly speaking, this result suggests that the Galileon theories~\eqref{cubic-galileon} may not be embedded into standard local UV complete QFTs. Then, let us regard the classicalization phenomenon as a mechanism for the UV completion of the Galileon theories. 
The cubic and quartic Galileon terms are schematically written as $\partial^{2k}\phi^n$ with $(k,n)=(2,3)$ and $(k,n)=(3,4)$, respectively. Both of them predict $\alpha=3/5$ from eq.~\eqref{alpha-classicaliz}, thus implying that the Galileon field is non-localizable with $\alpha=3/5$ in $D=4$ dimensions.

\subsubsection{Little String Theory}

Another possible candidate for non-localizable QFTs are Little String Theories~\cite{Seiberg:1997zk} which describe string theory or M-theory in a specific decoupling limit in which gravitational and other bulk degrees of freedom are decoupled from the fivebranes. The (non-gravitational) degrees of freedom living on the fivebrane are interacting and, because of the involved decoupling limit, the corresponding six-dimensional theory turns out to be different from the conventional local QFTs. The decoupling limit is defined by taking $g_s\rightarrow 0$ and $M_p\rightarrow \infty,$ while keeping the string scale $M_s$ fixed~\cite{Seiberg:1997zk}. This means that  string effects are still present and non-negligible although the string coupling $g_s$ and the gravitational coupling $1/M_p$ vanish. Indeed, the name Little String Theory derives from the fact that some of the properties of string theory that are not based on standard local QFT are still mantained; e.g., $T$-duality is still respected.

It has been claimed that Little String Theories can be described as quasi-local QFTs (i.e. with $\alpha=1/2$) because the Wightman function of the corresponding fields can grow exponentially as $\rho(\mu)\sim e^{c\sqrt{\mu}}$~\cite{Aharony:1998tt,Kapustin:1999ci}. In particular, this exponential growth was argued based on the Hagedorn behavior of the density of energy states $\rho(E)\sim e^{c\,E},$ where $E=\sqrt{\mu}.$

Before concluding this section, it is worth mentioning that even in \textit{perturbative string theory} non-local features due to string effects naturally emerge~\cite{Eliezer:1989cr}. 
The non-local nature of perturbative string theory has been studied by investigating the high-energy behavior of the scattering between strings~\cite{Gross:1987kza,Gross:1987ar,Mende:1989wt}; see sec.~\ref{sec:examp-string} for more discussions.

\section{Basics \& assumptions}\label{sec:assump}

In the next two sections we are going to prove \textit{two} lower bounds on how fast an elastic scattering amplitude can decrease in \textit{two} different kinematic regimes: in the hard-scattering limit (high center-of-mass energy and fixed angle) and in the Regge limit (high center-of-mass energy and fixed momentum transfer).  Unlike the proof of the Cerulus-Martin bound~\cite{Cerulus:1964cjb}, more generally we will assume exponential boundedness instead of polynomial boundedness. To derive such bounds several mathematical details  need to be explained. In this section we 
are going to introduce some basics of kinematics and discuss the assumptions used for the proofs in the next sections.

\subsection{Basics of kinematics} \label{sec:setup}

Let us decompose the S-matrix $\bm{S}$ into an identity matrix $\bm{1}$ and the transfer matrix $\bm{T}$ as $\bm{S}= \bm{1}+i\bm{T}$. We define the scattering amplitude $\scat$ from the initial state $\ket{i}$ to the final state $\ket{f}$ as 
\begin{align}
    \bra{f} \bm{T} \ket{i}
    =
    (2\pi)^4\delta^{(4)}(p_i-p_f) \scat(i\to f)
    \,,
\end{align}
where $p_i$ and $p_f$ are momentum eigenvalues of $\ket{i}$ and $\ket{f}$, respectively.
We now consider a $2\rightarrow 2$ elastic scattering with incoming momenta $p_1,p_2$ and outgoing momenta $p_3,p_4,$ such that $p_1+p_2=p_3+p_4$. The corresponding scattering amplitude $\scat$ is defined by 
\begin{align}
    \langle p_3,p_4| \bm{T} |p_1,p_2\rangle
    =
    (2\pi)^4\delta(p_1+p_2-p_3-p_4)\scat (s,t,u)
    \,,
\end{align}
where $s$, $t$, and $u$ are the Mandelstam variables, 
\begin{eqnarray}
s=-(p_1+p_2)^2\,,\quad t=-(p_1-p_3)^2\,, \quad u=-(p_1-p_4)^2\,.
\end{eqnarray}
For simplicity we assume that all masses are equal, $p_i^2=-m^2,$ thus the relation $s+t+u=4m^2$ holds. By working in the center-of-mass frame, i.e., $\vec{p}_1+\vec{p}_2=\vec{0}=\vec{p}_3+\vec{p}_4,$ and defining the scattering angle $\theta$ through the relation $\vec{p}_1\cdot\vec{p}_3=|\vec{p}_1||\vec{p}_3|\cos\theta,$ the variable $s$ coincides with the center-of-mass energy squared, $s=(p^0_1+p_2^0)^2=(p^0_3+p_4^0)^2,$ whereas the other two Mandelstam variables can be written as
\begin{equation}
t=-2p^2_s(1-\cos\theta)\,,\quad u=-2p_s^2(1+\cos\theta)\,,
\end{equation}
where $p_s$ is the modulus of the spatial momentum in the center-of-mass frame $p_s=|\vec{p}_1|=|\vec{p}_3|=\frac{1}{2}\sqrt{s-4m^2}.$ The variables $t$ and $u$ have the physical meaning of momentum transfer squared.

Because of the condition $s+t+u=4m^2$, we can regard $\scat(s,t,u)$ as a function of two kinematic invariants. In this paper we will consider two different choices: we will work either with the variables $s$ and $z=\cos\theta=1+t/2p^2_s,$ or with $s$ and $t.$  In the former case we write $\scat$ as $\scat(s,z);$ whereas by using an abuse of notation we write $\scat(s,t)$ in the latter case. These two functions are related as $\scat(s,z)=\scat(s,t(z))$ with $t(z)=-2p_s^2(1-z)$. We use $\scat(s,z)$ when we analyze the amplitude in the high-energy limit $s\to\infty$ with fixed $z$. This includes the hard-scattering regime defined as 
\begin{equation}
\text{hard-scattering regime:}\qquad s\rightarrow\infty\,,\quad -1<\cos\theta<1 ={\rm fixed}\,. 
\end{equation}
Whereas, we use $\mathcal{M}(s,t)$ when we consider the Regge regime:
\begin{equation}
\text{Regge regime:}\qquad s\rightarrow\infty\,,\quad t ={\rm fixed}\,. 
\end{equation}
This limit includes the high-energy scattering processes at small angles $|\theta|\ll1$. For large but finite $s$, we define the Regge limit as $s\gg m^2, |t|$, whereas the hard-scattering limit is $s\sim-t\gg m^2$.

We assume that the amplitude $\scat(s,z)$ is analytic in the complex $z$-plane for $s\geq 4m^2$, except for the presence of usual branch cuts on the real axis for $s\geq 4m^2,$ $z\geq 1+2m^2/p_s^2$ and $z\leq -1-2m^2/p_s^2.$ These details will be important for the next subsection.

\subsection{Analyticity domain in the $\cos \theta$-plane} 

Let us now introduce some mathematical details that are needed for both derivations in secs.~\ref{sec:lower-bound} and~\ref{sec:lower-bound-Regge}. 
We define a subdomain $\mathcal{D}_R$ in a finite region of the $z$-cut plane.
Though the precise shape of $\mathcal{D}_R$ is given in the paper by Cerulus and Martin~\cite{Cerulus:1964cjb}, 
here we will proceed in several steps and provide more details as its definition is not straightforward.
To define it we first make two variable transformations $z\rightarrow w\rightarrow \tau,$ and by imposing some conditions in the complex $\tau$-plane we fix the shape of $\mathcal{D}_R.$ Let us show how this can be done.

As we mentioned in sec.~\ref{sec:setup}, $\scat(s,z)$ has two branch cuts in the $z$-plane which run from $-\infty$ to $-\rho$ and from $\rho$ to $+\infty,$ respectively; we defined $\rho\coloneqq 1+2m^2/p_s^2$. Let us also introduce two points $z=-a$ and $z=a$ on the real axis such that $0< a<1.$ Then, we have $\rho>1>a.$ Now we make the first transformation:
\begin{equation}
w(z)=\frac{\rho}{z}\left(\rho-\sqrt{\rho^2-z^2}\right)\,,\quad w\in\mathbb{C}\,,\label{eq:wz}
\end{equation}
and define
\begin{eqnarray}
&&w(z=\pm 1)=\pm\rho\left(\rho-\sqrt{\rho^2-1}\right)\equiv \pm \varepsilon\,,\\[2mm]
&&w(z=\pm a)=\pm\frac{\rho}{a}\left(\rho-\sqrt{\rho^2-a^2}\right)\equiv \pm a'\,.
\end{eqnarray}
From these definitions, it is easy to verify the inequalities $\rho>1>\varepsilon>a'>0$.

We now make a second transformation from $w$ to a new variable $\tau$ that is defined such that the segment $[-a',a']$ in the $w$-plane is mapped into a unit circle in the $\tau$-plane. This map is given by
\begin{eqnarray}
\tau(w)=\frac{1}{a'}\left(w+\sqrt{w^2-a'^2}\right)\,,\quad \tau\in\mathbb{C}\,.\label{tau-variable}
\end{eqnarray}
Let us check that the segment $w\in [-a',a']$, i.e., the domain $\mathcal{W}_\text{sg}=\mathcal{W}_\text{sg}^+\cup \mathcal{W}_\text{sg}^-$ where $\mathcal{W}_\text{sg}^\pm\equiv\{w\in\mathbb{C}:\,-a'<\mathrm{Re}\,w< a'\,,\,\mathrm{Im}\,w=\pm0^+ \},$ is mapped into the unit circle in the $\tau$-plane. For $w\in \mathcal{W}_\text{sg}^\pm$, we have
\begin{align}
    \sqrt{w^2-a'^2}=\sqrt{a'^2-w^2}e^{\pm i\pi/2}=\pm i \sqrt{a'^2-w^2}\,,\label{sqrtcontinuation}
\end{align}
and by writing $\re\,w=a'\cos\vartheta$ we obtain a simple expression for $\tau(w):$ 
\begin{align}
    \tau(w)=\cos\vartheta\pm i\sin\vartheta=e^{\pm i\vartheta}\qquad (w\in \mathcal{W}_\text{sg})\,.
\end{align}
This concludes that the domain $\mathcal{W}_\text{sg}$ in the $w$-plane is mapped to the unit circle in the complex $\tau$-plane.

We now define
\begin{eqnarray}
&&\tau(w=\rho)=\frac{1}{a'}\left(\rho+\sqrt{\rho^2-a'^2}\right)\equiv R\,,\\[2mm]
&&\tau(w=\varepsilon)=\frac{1}{a'}\left(\varepsilon+\sqrt{\varepsilon^2-a'^2}\right)\equiv E\,.
\end{eqnarray}
Since $\rho>1>\varepsilon>a',$ it follows that $R>E>1.$

\begin{figure}[t!]
	\centering
		\includegraphics[scale=0.41]{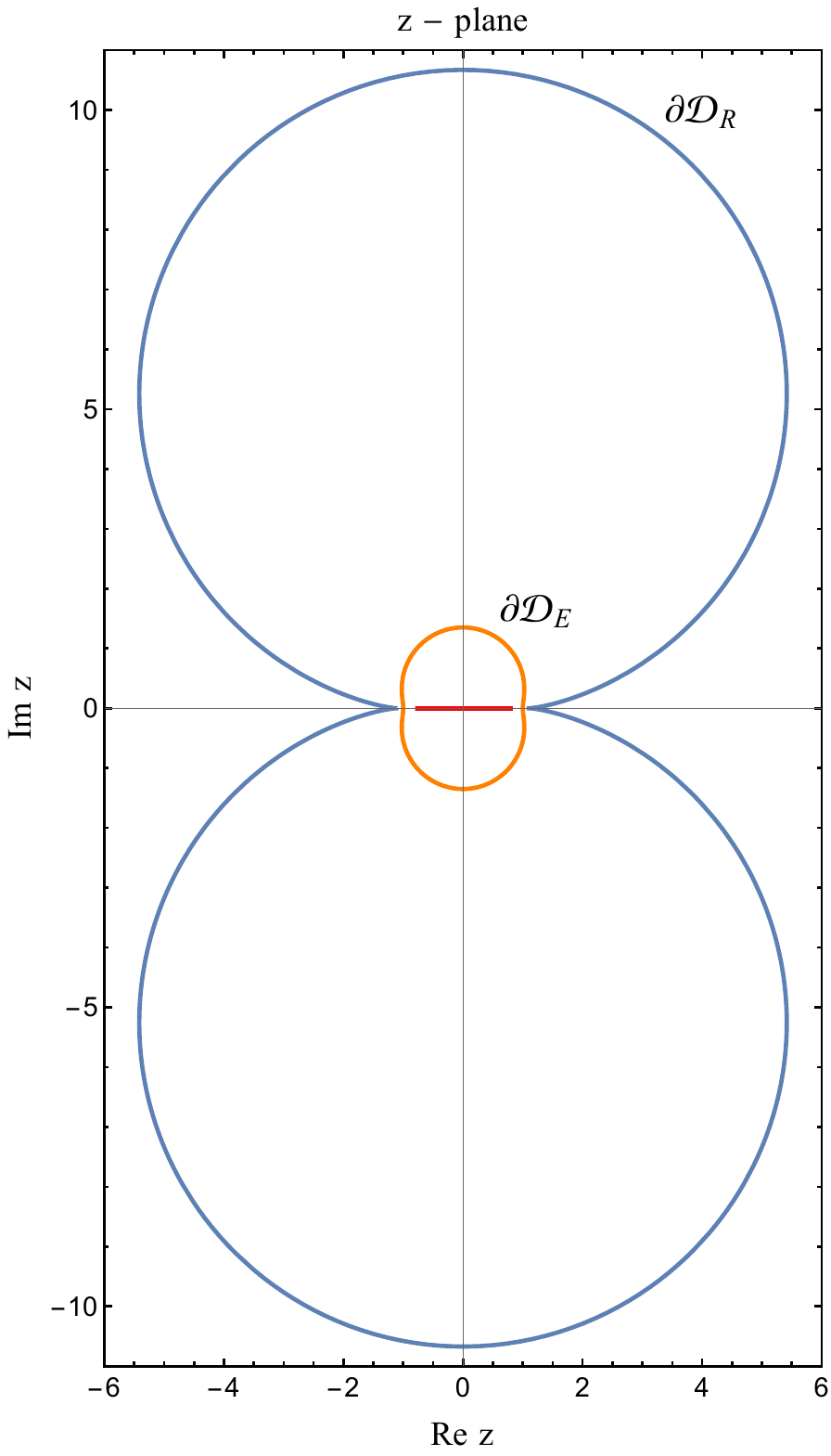}
	\protect\caption{This figure shows the boundary domain $\partial\mathcal{D}_R$ (blue line) in the $z$-plane defined in eq.~\eqref{def-D}. Below we show that $\partial\mathcal{D}_R$ is mapped to the ellipse $\mathcal{W}_{\beta_R}$ in the $w$-plane and to the circle of radius $R$ in the $\tau$-plane. Moreover, the red line corresponds to the segment $[-a,a]$, while the orange line to the boundary domain $\partial \mathcal{D}_E$ that is mapped to the ellipse $\mathcal{W}_{\beta_E}$ in the $w$-plane and to the circle of radius $E$ in the $\tau$-plane as we will show below.  To draw this figure we set $\rho=1.1$ and $a=0.8,$ which also give $a^\prime\simeq 0.474$.}\label{fig1}
\end{figure}

We choose the subdomain $\mathcal{D}_R$ in the $z$-plane by requiring that its boundary corresponds to the circle of radius $R$ in the $\tau$-plane, i.e.,
\begin{eqnarray}
\partial \mathcal{D}_R=\left\lbrace z\in \mathbb{C}:\,|\tau(w(z))|^2=R^2 \right\rbrace\,.\label{def-D} 
\end{eqnarray}
The shape of $\partial\mathcal{D}_R$ is the dumb-bell shown in Fig.~\ref{fig1}, and we will explicitly determine it below.


\begin{figure}[t!]
	\centering
	\subfloat[Subfigure 1 list of figures text][]{
		\includegraphics[scale=0.425]{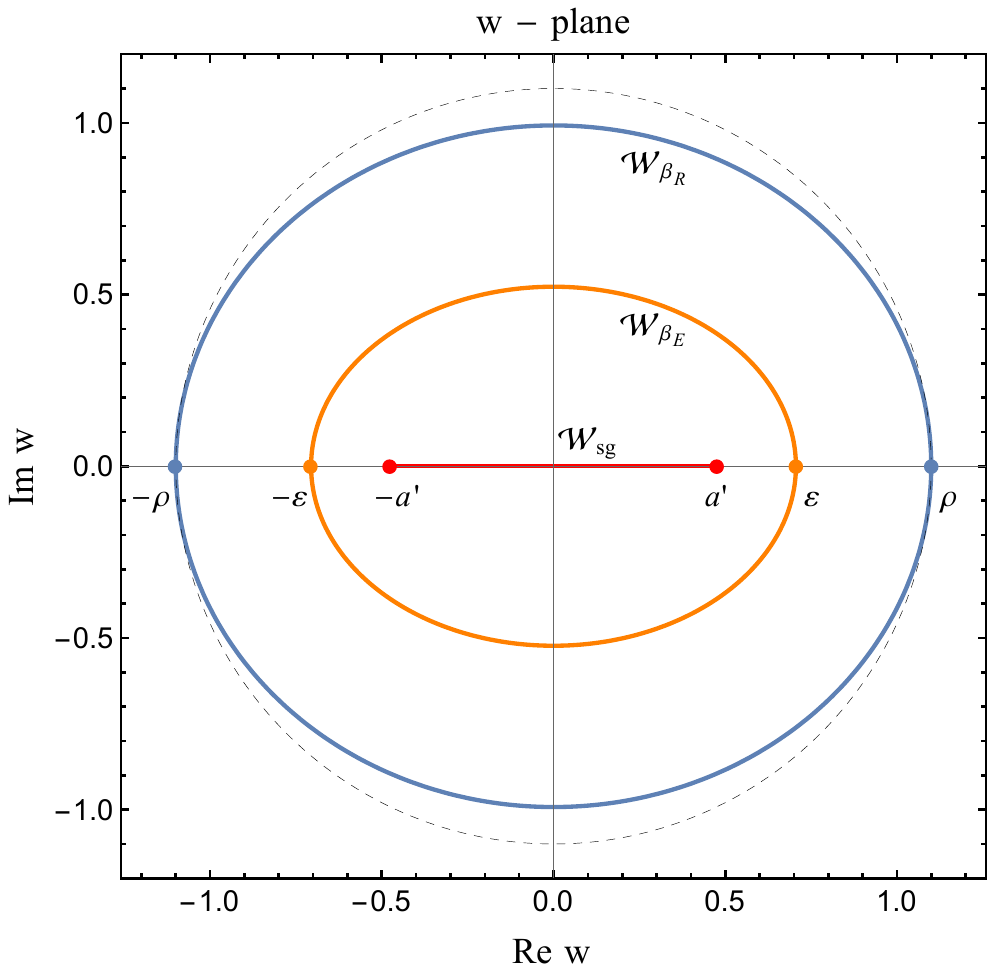}\label{fig2a}}\qquad\,\,
	\subfloat[Subfigure 2 list of figures text][]{
		\includegraphics[scale=0.41]{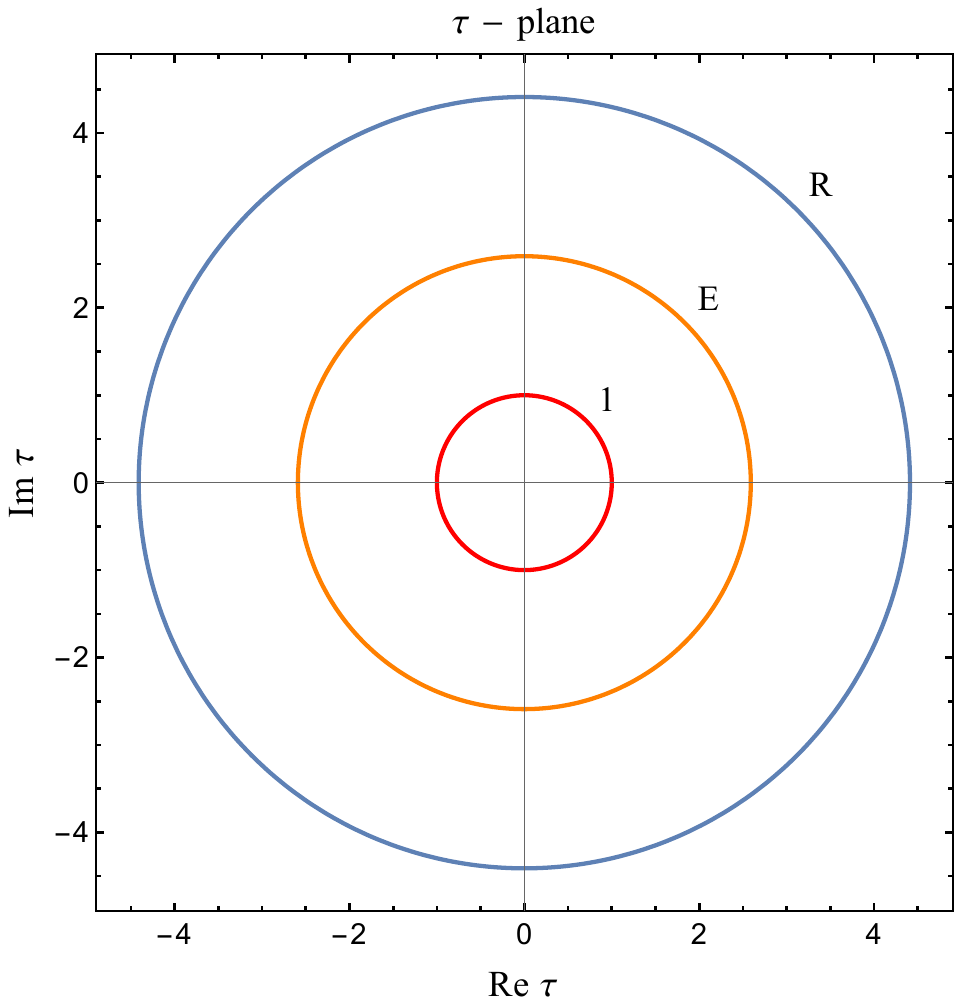}\label{fig2b}}
	\protect\caption{(a) Ellipses $\mathcal{W}_{\beta_{R}}$ (bue line), $\mathcal{W}_{\beta_{E}}$ (orange line) and segment $\mathcal{W}_{sg}$ (red line) in the $w$-plane. The ellipse $\mathcal{W}_{\beta_{R}}$ is tangent to the circumference of radius $\rho$ (dashed black line).  (b) The respective corresponding three circles of radius $|\tau|=R$ (blue line), $|\tau|=E$ (orange line) and $|\tau|=1$ (red line) in the $\tau$-plane. To draw these figures we set $\rho=1.1$ and $a=0.8,$ which also give $a^\prime\simeq 0.474$ and $\varepsilon \simeq 0.7$.}\label{fig2}
\end{figure}

\paragraph{From $w$-plane to $\tau$-plane.}
We now clarify how the domains in the $w$-plane look like in the $\tau$-plane. For this purpose, we introduce two real variables $\beta$ and $\varphi$ which are related to $w$ as $w=a'\cosh(\beta+i\varphi)$. 
With this parametrization, the domain $\mathcal{W}_\beta$
defined as 
\begin{equation}
\mathcal{W}_\beta\equiv\{w\in\mathbb{C}: w=a'\cosh(\beta+i\varphi),\,-\pi<\varphi\leq\pi\}
\end{equation}
is an ellipse\footnote{This can be easily checked by writing $w=a'\cosh(\beta+i\varphi)=x+iy$, with $x=a'\cosh\beta\cos\varphi$ and $y=a'\sinh\beta\sin\varphi,$ from which one can derive the ellipse equation $x^2/(a'\cosh\beta)^2+y^2/(a'\sinh\beta)^2=1$.} with foci $w=\pm a',$ semi-major axis $a'\cosh \beta$ and semi-minor axis $a'\sinh \beta$. In the limit $\beta\to0^+$, the ellipse $\mathcal{W}_\beta$ shrinks and coincides with the domain $\mathcal{W}_\text{sg}$ because we have $\im\,w\to0^+\sin\varphi$ and $\re\,w\to a'\cos\varphi$. In terms of the variables $\beta$ and $\varphi$, we have a simple expression for $\tau(w)$ of the form 
\begin{align}
    \tau=e^{\beta+i\varphi}\,.
\end{align}
This relation makes it manifest that the domain $\mathcal{W}_\beta$ in the $w$-plane is mapped to the circle with radius $e^\beta$ in the $\tau$-plane. In the limit $\beta\to0^+$, this result correctly reproduces the statement that the domain $\mathcal{W}_\text{sg}$ in the $w$-plane is mapped to the unit circle in the $\tau$-plane. In particular, we define $\beta_E$ and $\beta_R$ by 
\begin{align}
    \cosh\beta_E=\frac{\varepsilon}{a'}\,,\qquad\cosh\beta_R=\frac{\rho}{a'}\,.
\end{align}
Note that we have $\beta_R>\beta_E>0$ as a consequence of the inequality $\rho>\varepsilon>a'$. Then, the ellipses $\mathcal{W}_{\beta_{R}}$ and $\mathcal{W}_{\beta_{E}}$ in the $w$-plane are mapped to the circles with radius $R$ and $E$ in the $\tau$-plane, respectively. The ellipse $\mathcal{W}_{\beta_R}$ is tangent to and lies within the circumference of radius $\rho$ in the $w$-plane. Because $|\tau|=e^\beta$ is a monotonically increasing function of $\beta$, we conclude that the domain inside the circle with radius $R$ is the image of the domain inside the ellipse $\mathcal{W}_{\beta_{R}}$ in the $w$-plane. In Fig.~\ref{fig2a} we showed the shapes of the domains $\mathcal{W}_\text{sg}$, $\mathcal{W}_{\beta_E}$ and $\mathcal{W}_{\beta_R}$ in the $w$-plane, and in Fig.~\ref{fig2b} the three corresponding circles of radius $|\tau|=1,E,R$ in the $\tau$-plane.

\paragraph{From $w$-plane to $z$-plane.}
Next, we investigate how the ellipses $\mathcal{W}_\beta$ look like in $z$-plane, i.e., we derive the shape of $\partial \mathcal{D}_R$. 
To do so, it is convenient to solve~\eqref{eq:wz} in terms of $z,$ thus we obtain
\begin{align}
    z(w)=\frac{2\rho^2w}{w^2+\rho^2}\,.\label{analytic:zw}
\end{align}
which is analytic except for the poles at $w=\pm i\rho$. 


The expression for $z(w)$ allows us to investigate how the ellipses $\mathcal{W}_\beta$ look like in $z$-plane by simply substituting $w=a'\cosh(\beta+i\varphi)$ into \eqref{analytic:zw}. By writing $w=a'\cosh(\beta+i\varphi)=x+iy$, with $x=a'\cosh\beta\cos\varphi$ and $y=a'\sinh\beta\sin\varphi,$ we obtain the following parametric equations:
\begin{eqnarray}
\re\, z(w)=\frac{2\rho^2x(x^2+y^2+\rho^2)}{(x^2-y^2+\rho^2)^2+4x^2y^2}\,,\qquad
\im\, z(w)= -\frac{2\rho^2y(y^2+x^2-\rho^2)}{(x^2-y^2+\rho^2)^2+4x^2y^2}\,.
\label{param-eqs}
\end{eqnarray}
In particular, we find that the ellipse $\mathcal{W}_{\beta_{R}}$ in the $w$-plane is the image of the dumb-bell shaped curve $\partial\mathcal{D}_R$ defined in  eq.~\eqref{def-D} and shown in Fig.~\ref{fig1}, which is parameterized by
\begin{align}
    \re\,z|_{\beta_R}=\frac{2\rho ^3 \cos \varphi \left(2 \rho ^2- a'^2 \sin ^2\varphi\right)}{a'^4 \sin ^4\varphi +4 \rho ^4 \cos ^2\varphi}\,,\qquad\im\,z|_{\beta_R}=\frac{2 \rho ^2 a'^2\sqrt{\rho ^2-a'^2} \sin ^3\varphi}{a'^4 \sin ^4\varphi +4 \rho ^4 \cos ^2\varphi }\,,\label{points}
\end{align}
with $-\pi<\varphi\leq\pi$. The parametric equations in eq.~\eqref{points} reproduce the curve in Fig.~\ref{fig1} for $\rho=1.1$ and $a=0.8.$ The intersection points of $\partial \mathcal{D}_R$ with the imaginary $z$-axis can be obtained from the condition $\re\,z|_{\beta_R}=0,$ i.e., by substituting $\varphi=\pm\pi/2$ into \eqref{points}. Writing these points as $z=\pm iz_0$, 
we have
\begin{align}
    z_0=\frac{2\rho^2\sqrt{\rho^2-a'^2}}{a'^2}
    \,.
    \label{inter section1}
\end{align}
As a consistencty check, we can also verify that for $\varphi=0$ and $\varphi=\pi$ we have $\re\,z|_{\beta_R}=\rho$ and $\re\,z|_{\beta_R}=-\rho,$ respectively.

Similarly, by evaluating $x$ and $y$ at $\beta=\beta_E$ in eq.~\eqref{param-eqs} we find that the ellipse $\mathcal{W}_{\beta_{E}}$ in the $w$-plane is the image of the dumb-bell shaped curve $\partial \mathcal{D}_E$ shown in Fig.~\ref{fig1}, where $\mathcal{D}_E$ is the domain bounded by $\partial \mathcal{D}_E.$ 
In addition, one can easily see that the segment $\mathcal{W}_{\rm sg}$ in the $w$-plane corresponds to the segment $[-a,a]$ in the $z$-plane with $a=2\rho^2a^\prime /(a^{\prime^2}+\rho^2).$

\subsection{Assumptions} \label{subsec:assumptions}

We now list the set of hypotheses that we assume in order to prove the lower bounds on the maximum of the modulus of the elastic scattering amplitude in the hard-scattering and Regge limits:
\begin{enumerate}
	
	\item $\scat(s,z)$ is analytic in the subdomain $\mathcal{D}_R$ of the cut $z$-plane.  \label{assump-1}
	
	\item  $\scat(s,z)$ is exponentially bounded on the boundary  $\partial \mathcal{D}_R$ as 
	\begin{equation}
	|\scat(s,z)|\leq A\left(\frac{s}{s_0}\right)^N\,e^{\sigma (s/s_0)^\alpha}\,,\qquad \frac{s}{s_0}\gg 1\,, 
 \label{assump-exp}
	\end{equation}
where $N$ and $\alpha$  are positive constants independent of $s$ and $z$, $\sigma$ and $A$ are positive parameters independent of $s$ but can in general depend on $z$, while $s_0$ denotes some energy-squared reference scale. \label{assump-2}

	\item The modulus of the forward scattering amplitude ($z=\cos\theta=1$) is bounded from below as 
	\begin{equation}
	\left|\mathcal{M}(s,1)\right|\geq  \frac{B}{(s/s_0)^\beta}\,,\qquad \frac{s}{s_0}\gg 1\,,  \label{bound-forward}
	\end{equation}
	where $\beta$ and $B$ are positive constants independent of $s$. The condition $\beta+N>0$ must be satisfied when $\alpha=0$. \label{assump-3}

\end{enumerate}
\paragraph{Remark.} Assumption~\ref{assump-1} 
was also assumed to prove the standard Cerulus-Martin bound in the hard-scattering regime~\cite{Cerulus:1964cjb}. Assumption~\ref{assump-2} is more general than the one used in Ref.~\cite{Cerulus:1964cjb} where polynomial boundedness was assumed (i.e., $\alpha=0$). Assumption~\ref{assump-3} was assumed in the initial work~\cite{Cerulus:1964cjb} by Cerulus and Martin, but then in Ref.~\cite{Jin:1964zz} it was proved that unitarity, dispersion relations and polynomial boundedness together imply the lower bound on the forward scattering amplitude: if we parameterize the forward scattering amplitude as $|\scat(s,1)|\sim s^{-\beta}$, then one can show eq.~\eqref{bound-forward} with $\beta\leq 2$~\cite{Jin:1964zz,Epstein:2019zdn}. Hence, the Cerulus-Martin bound can be essentially derived by replacing assumption~\ref{assump-3} with unitarity.
It is however still unclear whether eq.~\eqref{bound-forward} can be derived from unitarity for general amplitudes satisfying exponential boundedness. Therefore, in what follows we take the inequality~\eqref{bound-forward} as a starting assumption to prove the lower bounds, without using unitarity, as done in the initial proof by Cerulus and Martin~\cite{Cerulus:1964cjb}.

\section{Lower bound in the hard-scattering regime}\label{sec:lower-bound}

\subsection{Proof}\label{sec:proof-hard}

We now prove the following statement:\footnote{In our derivation we assume the presence of a finite mass gap as originally done by Cerulus and Martin in their proof~\cite{Cerulus:1964cjb}. However, it is worth mentioning that the case of vanishing mass gap can be taken into account by suitably modifying the definitions of the domains in the complex $z$-plane, for example by changing the location of $D_E$ and thus modifying assumption~\ref{assump-3} away from the forward limit. For instance, this has been done in Ref.~\cite{Tourkine:2023xtu} at least for the Cerulus-Martin bound, i.e., for gapless theories with $\alpha=0$. We expect that the only changes in the formula of the lower bound are different expressions for the functions $f(a)$ and $g(a).$
See the end of sec.~\ref{sec:discuss} for further discussion of the gapless case.
\label{foot-gapless}}

\medskip

\textit{If the assumptions~\ref{assump-1},~\ref{assump-2},
and~\ref{assump-3} on the elastic scattering amplitude $\mathcal{M}(s,\cos\theta)$ are satisfied, then in the presence of a finite mass gap it follows that in the high-energy limit and for fixed scattering angle (hard-scattering regime) the maximum of the modulus $\left|\mathcal{M}(s,\cos\theta)\right|$ is bounded as}
\begin{equation}
\boxed{\max_{-a\leq \cos\theta\leq a}\left|\mathcal{M}(s,\cos\theta)\right|\geq \mathcal{N}(s)\, e^{-f(a)\,\sqrt{s}\,\log (s/s_0)} e^{-g(a)\,s^{\alpha+\frac{1}{2}}}\,,
}\label{new-bound} 
\end{equation}
\textit{where $\mathcal{N}(s)$ is a positive function of $s$ that is subdominant in the $s\rightarrow\infty$ limit, and $f(a)$ and $g(a)$ are some positive functions of $a\in (0,1).$
}

\begin{proof}
	
Let us consider the three circles of radii $1,$ $E$ and $R$ in the $\tau$-plane (see Fig.~\ref{fig2b}). From the remarks made in the previous section we know that $1<E<R$, therefore the circle $|\tau|^2=R^2$ is the outer one, $|\tau|^2=E^2$ the intermediate, and $|\tau|^2=1$ the inner. 
Let us call the maxima of $\left|\scat(s,z)\right|$ on the three circles as
\begin{eqnarray}
M_R\equiv \max\limits_{z\in \partial\mathcal{D}_R} |\scat(s,z)|\,,\quad M_E\equiv \max\limits_{z\in \partial \mathcal{D}_E} |\scat(s,z)|\,,\quad  M_{1}\equiv\max\limits_{-a\leq z\leq a} |\scat(s,z)|\,,
\end{eqnarray}
respectively. 

Since $\scat(s,z)$ is analytic  in $\mathcal{D}_R$ (assumption~\ref{assump-1}), it will also be analytic in the annulus $1\leq |\tau|\leq R.$ Thus,  we can apply the Hadamard-three-circle theorem~\cite{hadamard-book} on the three circles in the $\tau$-plane, and obtain the following inequality:
\begin{eqnarray}
&&\log R\log M_E\leq \log\frac{R}{E}\log M_1+\log E\log M_R\nonumber \\[2mm]
\Leftrightarrow&& \log M_E\leq \left(1-\frac{\log E}{\log R}\right)\log M_1+\frac{\log E}{\log R}\log M_R\,.\label{hadamard}
\end{eqnarray}

We are interested in the high-energy behavior, i.e., in the large $s$ limit for which $\rho\rightarrow 1,$ $R\to E$ and $\log E/\log R\to1,$ thus it is convenient to introduce a positive function $C(x,a)$ defined by the following equation:
\begin{eqnarray}
\frac{\log E}{\log R}= 1-\frac{2m}{p_s}C(x,a)\,.
\end{eqnarray}
The explicit expression of $C$ is 
\begin{align}
    C(x,a)&=\frac{1}{\sqrt{x}}\frac{\log \left(\frac{\sqrt{2 (x+2) \left(\sqrt{(x+2)^2-4 a^2}-x-2\right)+8 a^2}+2 a}{\sqrt{2} \sqrt{(x+2) \left(a^2 \left(x-\sqrt{x (x+4)}+2\right)+\sqrt{(x+2)^2-4 a^2}-x-2\right)}+a \left(x-\sqrt{x (x+4)}+2\right)}\right)}{\log \left(\frac{\sqrt{2 (x+2) \left(\sqrt{(x+2)^2-4 a^2}-x-2\right)+8 a^2}+2 a}{x+2-\sqrt{(x+2)^2-4 a^2}}\right)}\,,
    \label{function-C(x)}
\end{align}
where $x\equiv (2m/p_s)^2,$ and its positivity follows from the inequality $R>E$. 

In terms of $C$, eq.~\eqref{hadamard} becomes
\begin{eqnarray}
\log M_E\leq \frac{2mC}{p_s}\log M_1+\left( 1-\frac{2mC}{p_s}\right)\log M_R\,,\label{hadamard-high-energy}
\end{eqnarray}
and exponentiating we get
\begin{eqnarray}
M_E\leq M_1^{2mC/p_s}\,M_R^{1-2mC/p_s}\,,\label{hadamard-exponentiating}
\end{eqnarray}
which can be written as
\begin{align}
  M_1^{2mC/p_s}\geq M_R^{-1+2mC/p_s}M_E\,.\label{hadamard-exp2}
\end{align}
Because the exponent of $M_R
$, i.e., the term $-1+2mC/p_s$, is always negative, the assumption~\ref{assump-2} gives a lower bound on the left-hand-side of~\eqref{hadamard-exp2}. We then get 
\begin{align}
  M_1\geq \exp\left\{\left(-\frac{p_s}{2mC}+1\right)\log\left[A(s/s_0)^Ne^{\sigma(s/s_0)^{\alpha}}\right]\right\}M_E^{p_s/2mC}\,.\label{assump-2-used}
\end{align}
Since $M_E\geq |\scat(s,z =1)|$, we notice that \eqref{assump-2-used} provides a lower bound on the high-energy behavior of the absolute value of the modulus of the amplitude ($M_1=\max_{-a\leq z\leq a}|\scat(s,z)|$) in terms of the forward-limit scattering amplitude. Then, by using the assumption~\ref{assump-3}, 
we obtain  
\begin{align}
 & \max_{-a\leq\cos\theta\leq a}|\scat(s,\cos\theta)| 
 \geq \exp\left[-F(s,a)\right]\,,\label{lowerbound}
\end{align}
where $F(s,a)$ is given by 
\begin{align}
    F(s,a)&\equiv \left(\frac{p_s}{2mC}-1\right)\log\left[A(s/s_0)^Ne^{\sigma(s/s_0)^{\alpha}}\right]-\frac{p_s}{2mC}\log\left[B(s/s_0)^{-\beta}\right]\nonumber\\[2mm]
&=  \frac{p_s}{2mC}\log\left[\frac{A}{B}\left(s/s_0\right)^{N+\beta}e^{\sigma (s/s_0)^\alpha} \right] - \log\left[A \left(s/s_0\right)^{N}e^{\sigma (s/s_0)^\alpha}\right] \,.
    \label{defF}
\end{align}
The inequality in eq.~\eqref{lowerbound} holds for large but finite values of $s$ for which the assumptions we imposed in sec.~\ref{subsec:assumptions} are satisfied. To understand the bound~\eqref{lowerbound} more clearly, we need the behavior of the function $C(x,a)$ in the $x\rightarrow 0$ limit. 
Its expansion around $x=0$ strongly depends on the type of kinematic regime under investigation. In this section, we are interested in the hard-scattering limit ($x\to 0$, $a=$ fixed), while the Regge limit will be discussed in the next section. Hence, we consider the expansion of $C(x,a)$ around $x=0$ by keeping $0<a<1$ fixed and obtain $C(x,a)=C_0(a)+ \mathcal{O}(\sqrt{x}),$ where\footnote{We can numerically confirm that this expansion works well for sufficiently small $x\lesssim 0.01$ unless $a$ is very close to unity to satisfy $1-x\lesssim a\leq 1$.}
\begin{align}
C_0(a)
&= \frac{a}{\sqrt{2(\sqrt{1-a^2}+a^2-1)}\left\lbrace \log\left[\frac{1}{a}(\sqrt{1-a^2}+1)\left(1+\frac{\sqrt{2}}{a}\sqrt{\sqrt{1-a^2}+a^2-1}\right) \right]\right\rbrace}\,.\label{C0def}
\end{align}
We can numerically verify the inequality $C(x,a)\geq C_0(a)$ for sufficiently small $x$ satisfying the condition $x\leq x_*(a)$ for given $a\in (0,1)$. Here, $x_*(a)\neq 0$ is defined by the equation $C(x_*(a),a)=C_0(a)$; see App.~\ref{app-C(x)} for more details. Since $x_*(a)$ is approximated as $x_*(a)\simeq 8(1-a)/9$ for $a\simeq 1$ (see eq.~\eqref{x_*(a)-approx_a1} in App.~\ref{app-C(x)}), the condition $x\leq x_*(a)$ is equivalent to $a\leq 1-\frac{9}{8}x$, approximately. 

Therefore, for $x\leq x_*(a)$ we have $C(x,a)\geq C_0(a)$ which implies
\begin{equation}
F(s,a)\leq  F(s,a)|_{C\rightarrow C_0(a)}\,,
\label{ineq-F(s,a)}
\end{equation}
where we have used 
\begin{equation}
\log\left[\frac{A}{B}\left(s/s_0\right)^{N+\beta}e^{\sigma (s/s_0)^\alpha} \right]>0\,,
\end{equation}
which is always valid at sufficiently high energies. Thus, from eq.~\eqref{lowerbound} we obtain
\begin{align}
 & \max_{-a\leq\cos\theta\leq a}|\scat(s,\cos\theta)| 
 \geq \exp\left[-F(s,a)|_{C\rightarrow C_0(a)}\right]\,,
 \label{lowerbound-C>C0}
\end{align}
or, equivalently,
\begin{align}
 & \max_{-a\leq\cos\theta\leq a}|\scat(s,\cos\theta)| 
 \geq \mathcal{N}(s)\,\exp\left\lbrace -\frac{p_s}{2mC_0(a)}\left[(N+\beta)\log\left(\frac{s}{s_0}\right) +\sigma\left(\frac{s}{s_0}\right)^\alpha \right] \right\rbrace \,,
 \label{lowerbound-2}
\end{align}
where 
\begin{align}
    \mathcal{N}(s)& \equiv A \,\left(\frac{B}{A}\right)^{\frac{p_s}{2mC_0(a)}} \left(\frac{s}{s_0}\right)^{N}e^{\sigma (s/s_0)^\alpha}
\end{align}
is a positive function of $s$ that is subdominant in the large $s$ limit.

Finally, by using\footnote{Note that, for $m\neq 0$ and for finite values of $s$ we have the strict inequality $p_s<\sqrt{s}/2.$ However, this inequality can be saturated in the strict infinite energy limit $s\rightarrow \infty.$ This is the reason why we wrote $p_s\leq \sqrt{s}/2.$} $p_s\leq\sqrt{s}/2,$ from the inequality~\eqref{lowerbound-2} we obtain the  lower bound
\begin{align}
 & \max_{-a\leq\cos\theta\leq a}|\scat(s,\cos\theta)| 
 \geq \mathcal{N}(s)\,e^{-f(a)\,\sqrt{s}\,\log(s/s_0)-g(a)\,s^{\alpha+\frac{1}{2}}}\,,
 \label{lowerbound-3}
\end{align}
which coincides with eq.~\eqref{new-bound}; we have defined $f(a)\equiv \frac{N+\beta}{4mC_0(a)}>0$ and $g(a)\equiv \frac{\sigma}{4mC_0(a) s_0^\alpha}>0.$ Thus, we have showed the existence of a lower bound on the maximum of the modulus of an elastic scattering amplitude in the hard-scattering regime.

\end{proof}

\paragraph{Remarks.}
We now summarize several remarks on our results.

\begin{itemize}

\item In the case $\alpha=0$ the scattering amplitude is polynomially bounded (see assumption~\ref{assump-2} above) and we consistently recover the Cerulus-Martin bound~\cite{Cerulus:1964cjb}, i.e., eq.~\eqref{new-bound} reduces to the inequality~\eqref{C-M-bound}. 
The lower bound in eq.~\eqref{lowerbound-C>C0},~\eqref{lowerbound-2}, or~\eqref{lowerbound-3} is valid for large but {\it finite} values of $s$ for which the condition $x\leq x_*(a)$ is satisfied thanks to the inequality $C(x,a)\geq C_0(a)$, rather than an approximation $C(x,a)\simeq C_0(a)$ which becomes exact only in the limit $s\to\infty$.
In particular, our result in eq.~\eqref{lowerbound-2} is more precise and even for $\alpha=0$ can be seen as an update of the original derivation of the Cerulus-Martin bound in~\cite{Cerulus:1964cjb}. 

\item The lower bound becomes trivial in the limit $a\to 0$ because we have $C(x,0)=C_0(0)=0$, which is also mentioned in~\cite{Kinoshita:1964bmc}. This can be understood intuitively as follows: the left-hand-side of~\eqref{lowerbound} becomes $|\scat(s,\cos\theta=0)|$ in this limit, and there is a possibility that $\scat(s,\cos\theta=0)$ exactly vanishes at least for some discrete set of points on the positive real $s$-axis.

\end{itemize}

\subsection{Discussion}

The important new result is that we have obtained a new lower bound in eq.~\eqref{lowerbound-C>C0},~\eqref{lowerbound-2}, or~\eqref{lowerbound-3} with $\alpha>0$ in the hard-scattering regime that applies to amplitudes for which polynomial boundedness is violated. The $\alpha$-dependent term is dominant at sufficiently high energies and the lower bound effectively reads  
\begin{equation}
\max_{-a\leq \cos\theta\leq a}\left|\mathcal{M}(s,\cos\theta)\right|\geq \tilde{\mathcal{N}}(s) e^{-g(a)\,s^{\alpha+\frac{1}{2}}}\qquad (\alpha>0)\,,
\label{bound-hard-non-local}
\end{equation}
where $\tilde{\mathcal{N}}(s)$ is a subdominant function of $s$ in the $s\rightarrow \infty$ limit. 
This shows that a violation of the Cerulus-Martin bound does not always signal a pathology because in the absence of polynomial boundedness a more general lower bound holds.

Our result admits a violation of the original Cerulus-Martin bound even when the theory is strictly localizable $0<\alpha<1/2$. This is interesting since the Cerulus-Martin bound has been used as a test of locality in the past.  
However, it may also be that there is a stronger bound than ours.
This is logically possible since our bound provides only a necessary condition. In fact, it would be interesting to understand whether one can find another way to prove the lower bound of the Cerulus-Martin type \eqref{C-M-bound} even for $0<\alpha<1/2$ by changing some of the assumptions; see also sec.~\ref{sec:examples} for more discussions on this point. We leave this aspect for future work.

\section{Lower bound in the Regge regime}\label{sec:lower-bound-Regge}

We now focus on the Regge limit ($s\rightarrow \infty,$ $t=-2p_s^2(1-\cos\theta)={\rm fixed}$) and show the existence of a lower bound in this regime for both cases of polynomial ($\alpha=0$) and exponential ($\alpha>0$) boundedness. To our knowledge, our lower bound on the amplitude in the Regge limit is new in the literature, even in the $\alpha=0$ case.

The main difference with the hard-scattering regime is that, in the Regge limit, we should not expand the function $C(x,a)$ around $x=0$ for fixed values of $a,$ because the variable that now has to be kept fixed is the momentum transfer squared $t.$ To correctly take into account the Regge limit, we introduce the quantity
\begin{equation}
\Delta\equiv\frac{1-a}{x}=\frac{p_s^2}{4m^2}(1-a)
\,,
\end{equation}
such that to derive a lower bound we now need to expand the function $C(x,\Delta)\equiv C(x,a)|_{a\rightarrow 1-x\Delta}$ around $x=0$ and for fixed values of $\Delta.$   In terms of $\Delta,$ the momentum transfer squared can be expressed as $t=-8m^2\Delta|_{a=\cos\theta}.$ Moreover, in terms of $t$ and $\Delta$ the inequalities $-a\leq \cos\theta\leq a$ become $8m^2\Delta-4p_s^2\leq t\leq -8m^2 \Delta.$ 

\subsection{Proof}\label{sec:proof-regge}

We now prove the following statement:

\medskip

\textit{If the assumptions~\ref{assump-1},~\ref{assump-2},
and~\ref{assump-3} on the elastic scattering amplitude $\mathcal{M}(s,t(\cos\theta))$ are satisfied, then in the presence of a finite mass gap it follows that in the high-energy limit and for fixed negative momentum transfer squared (Regge regime) the maximum of the modulus $\left|\mathcal{M}(s,t)\right|$ is bounded as}
\begin{equation}
\boxed{\max_{8m^2\Delta-4p_s^2\leq t\leq -8m^2 \Delta}\left|\mathcal{M}(s,t)\right|\geq h(\Delta) \, e^{-\tilde{f}(\Delta) \log(s/s_0)-\tilde{g}(\Delta)s^\alpha}\,,
}\label{new-bound-regge} 
\end{equation}
\textit{where $h(\Delta),$ $\tilde{f}(\Delta)$ and $\tilde{g}(\Delta)$ are positive functions of $0<\Delta\leq p_s^2/4m^2$ defined in \eqref{func-def-regge}.}

\begin{proof}

The starting point is eq.~\eqref{lowerbound} rewritten for fixed momentum transfer $t<0,$ i.e.,
\begin{align}
 & \max_{8m^2\Delta-4p_s^2\leq t\leq -8m^2 \Delta}|\scat(s,t)| 
 \geq \exp\left[-F(s,\Delta)\right]\,,\label{lowerbound-fix-t}
\end{align}
where $F(s,\Delta)\equiv F(s,a)|_{a\rightarrow 1-x\Delta},$ and $F(s,a)$ was defined in eq.~\eqref{defF}. Because we are now interested in the Regge limit, we expand the function $C(x,\Delta)\equiv C(x,a)|_{a\to1-x\Delta}$ around $x=0$ by keeping $\Delta$ fixed as\footnote{We note that some details on the behavior of the function $C(x,a)$ in the Regge limit (i.e., of $C(x,\Delta)$ in the limit $x\rightarrow 0$ and for fixed $\Delta$) are also given in the footnote 11 of~\cite{PhysRevLett.12.257}.}
\begin{align}
    C(x,\Delta)=\left[1-6C_0(\Delta)\right]\frac{1}{\sqrt{x}}+C_0(\Delta)+\mathcal{O}\left(\sqrt{x}\right)\,,
\end{align}
where
\begin{equation}
C_0(\Delta)=\frac{\sqrt{\sqrt{2 \Delta +1}-1}}{6 ({2 \Delta +1})^{1/4}}\,,\label{CDelta-def}
\end{equation}
and we have $1-6C_0(\Delta)>0$ for $\Delta>0.$ We can notice that the small $x$ behavior of the function $C(x,\Delta)$ in the Regge regime (i.e., small angles) is qualitatively different from the one in the hard-scattering regime (i.e., large angles) derived in the previous section. 

We can numerically verify the inequality $C(x,\Delta)\geq [1-6C_0(\Delta)]x^{-1/2}$ for sufficiently small $x$ satisfying the condition $x\leq x_*(\Delta)$. Here, $x_*(\Delta)\neq0$ is defined by the equation $C(x_*(\Delta),\Delta)=[1-6C_0(\Delta)]x_*^{-1/2}$; see App.~\ref{app-C(x)} and Fig.~\ref{figC-delta} for more details. Therefore, for $x\leq x_*(\Delta)$ we can write 
\begin{equation}
F(s,\Delta)\leq  F(s,\Delta)|_{C\rightarrow [1-6C_0(\Delta)]x^{-1/2}}\,.
\end{equation}
Thus, from eq.~\eqref{lowerbound-fix-t} we obtain
\begin{align}
 & \max_{8m^2\Delta-4p_s^2\leq t\leq -8m^2 \Delta}|\scat(s,t)|
 \geq \exp\left[-F(s,\Delta)|_{C\rightarrow [1-6C_0(\Delta)]x^{-1/2}}\right]\,,
 \label{lowerbound-2-fix-t}
\end{align}
or, equivalently,
\begin{align}
 & \max_{8m^2\Delta-4p_s^2\leq t\leq -8m^2 \Delta}\left|\mathcal{M}(s,t)\right|\geq h(\Delta) \, e^{-\tilde{f}(\Delta) \log(s/s_0)-\tilde{g}(\Delta)s^\alpha}\,,
 \label{lowerbound-2-regge}
\end{align}
where we have defined
\begin{equation}
h(\Delta)\equiv \left(\frac{B}{A^{6C_0(\Delta)}}\right)^{\frac{1}{1-6C_0(\Delta)}}\,,\quad \tilde{f}(\Delta)\equiv \frac{6C_0(\Delta)N+\beta}{1-6C_0(\Delta)}\,,\quad \tilde{g}(\Delta)\equiv \frac{6C_0(\Delta)}{1-6C_0(\Delta)}\frac{\sigma}{s_0^\alpha}\,,
\label{func-def-regge}
\end{equation}
which are positive functions of $\Delta$.

\end{proof}

\paragraph{Remarks.} We now summarize several remarks on our results.

\begin{itemize}
\item In the forward limit $t=0$ we have $\Delta=0$ (i.e., $a=1$), $C_0(\Delta=0)=0$ and $C(x,\Delta=0)=1/\sqrt{x}.$ As a consequence, we obtain $h(0)=B,$ $\tilde{f}(0)=\beta,$ $\tilde{g}(0)=0,$ and 
\begin{equation}
|\mathcal{M}(s,t=0)|=|\mathcal{M}(s,\cos\theta=1)|\geq \frac{B}{(s/s_0)^\beta}\,,
\end{equation}
which, as a consistency check, coincides with the assumption~\ref{assump-3}. The importance of the newly derived inequality~\eqref{new-bound-regge} is that it provides a lower bound for fixed momentum transfer and beyond the forward limit (i.e., 
 for $t< 0$ and $\Delta>0$).

\item The $s$-dependence in the lower bound for fixed momentum transfer differs from the one for fixed scattering angle by a factor of $\sqrt{s}$ in the exponent. This means that  amplitudes in the hard-scattering regime (large angles) can be more suppressed as compared to the ones in the Regge regime (small angles). This could be understood physically by observing that the probability of two particles scattering should be greater for smaller scattering angles.

\end{itemize}

\subsection{Discussion}

We have derived a new bound lower bound~\eqref{new-bound-regge} on elastic scattering amplitudes in the Regge limit. To our knowledge, this result is new in the literature even in the case of polynomial boundedness $\alpha=0$; 
the only work that is close to our analysis in this section is Ref.~\cite{PhysRevLett.12.257} where some details on the Regge limit of the function $C(x,\Delta)$ were discussed  but no bound was derived.

In the case $\alpha=0$, the lower bound~\eqref{new-bound-regge} reduces to
\begin{eqnarray}
\max_{8m^2\Delta-4p_s^2\leq t\leq -8m^2 \Delta}\left|\mathcal{M}(s,t)\right|\geq \tilde{h}(\Delta) \, \left(\frac{s_0}{s}\right)^{\tilde{f}(\Delta)} \qquad (\alpha=0)\,,
 \label{lowerbound-regge-alpha=0}
\end{eqnarray}
where $\tilde{h}(\Delta)\equiv h(\Delta)e^{-\tilde{g}(\Delta)}.$ Thus, under the assumption of locality ($\alpha=0$), in the Regge limit the elastic scattering amplitude is bounded from below by inverse powers of $s$. 

In the case $\alpha>0,$ the inequality~\eqref{new-bound-regge} reduces to
\begin{align}
 & \max_{8m^2\Delta-4p_s^2\leq t\leq -8m^2 \Delta}\left|\mathcal{M}(s,t)\right|\geq H(s,\Delta) \, e^{-\tilde{g}(\Delta)s^\alpha}\qquad (\alpha>0)\,,
 \label{lowerbound-2-regge-alpha>0}
\end{align}
where $H(s,\Delta)\equiv h(\Delta)e^{-\tilde{f}(\Delta) \log(s/s_0)}$ is a positive function that is subdominant in the $s\rightarrow \infty$ limit. Therefore, unlike the case $\alpha=0,$ for $\alpha>0$ the elastic scattering amplitude in the Regge limit turns out to be exponentially bounded from below.

We can also ask whether there is a regime where the bound in the Regge limit overlaps with the one in the the hard-scattering limit. This actually happens when $\Delta\gg 1$ which allows to expand $\tilde{f}(\Delta)=2(N+\beta)\sqrt{2\Delta}-(3N+\beta)/2+\mathcal{O}(\Delta^{-1/2})$ and $\tilde{g}(\Delta)=\sigma[2\sqrt{2\Delta}-3/2+\mathcal{O}(\Delta^{-1/2})]/s_0^\alpha.$ Thus, substituting these expansions in eq.~\eqref{new-bound-regge} and using $\sqrt{\Delta}\sim \sqrt{s}$ we get the same $s$-dependence in the exponent as in eq.~\eqref{new-bound}.

\subsection{Bound on the Regge trajectory}

As an application of the newly derived bound~\eqref{new-bound-regge}, it is interesting to constrain the Regge trajectory $j(t)$ in the regime $-s\ll t\leq0$.

Suppose that the amplitude in the Regge regime is dominated by the contribution from the leading Regge pole. In such a case we can parametrize the behavior of the amplitude in the regime $-s/t\gg 1$ as~\cite{Collins:1977jy,White:2019ggo}
\begin{align}
    \scat(s,t)= B(s,t) \left(\frac{s}{s_0}\right)^{j(t)}\,, \label{reggebehave}
\end{align}
where $B(s,t)$ is a subdominant multiplicative factor, and $j(t)$ is the so-called \textit{Regge trajectory} which, in general, is real for $t\in \mathbb{R}$ but can take complex values for $t\in \mathbb{C}$. 
Suppose that the amplitude~\eqref{reggebehave} satisfies the requirement of local QFT and is polynomially bounded on the domain $\der\mathcal{D}_R$ in the complex $z$-plane. Then it has to satisfy the lower bound with $\alpha=0$ in eq.~\eqref{lowerbound-regge-alpha=0}: 
\begin{equation}
\max_{8m^2\Delta-4p_s^2\leq t\leq -8m^2\Delta}\left[B(s,t)\left(\frac{s}{s_0} \right)^{j(t)}\right]\geq \tilde{h}(\Delta)\left(\frac{s_0}{s}\right)^{\tilde{f}(\Delta)}\,,\label{bound-regge-leading}
\end{equation}
where $j(t)$ takes only real values as the maximum is taken for negative real values of $t.$  If $j(t)>0,$ the lower bound is always satisfied because $\tilde{f}(\Delta)>0.$ The interesting case is when the Regge trajectory is negative. Indeed, for $j(t)<0$ the inequality~\eqref{bound-regge-leading} can be recast as a non-trivial bound on the Regge trajectory:\footnote{Note that, the inequality~\eqref{bound-regge-traject-1} can be saturated only in the strict $s\rightarrow \infty$ limit, in which subdominant $(s,t)$-dependent terms, such as $\log B(s,t)/\log (s/s_0)$ and $\log \tilde{h}(\Delta)/\log (s/s_0),$ become completely negligible on both sides of the inequality.}
\begin{equation}
\max_{8m^2\Delta-4p_s^2\leq t\leq -8m^2\Delta} j(t)\geq-\tilde{f}(\Delta)=-\frac{6C_0(\Delta)N+\beta}{1-6C_0(\Delta)}\,.\label{bound-regge-traject-1}
\end{equation}
Since the amplitude~\eqref{reggebehave} is a polynomial in $s$ we can easily link $\beta>0$ and $N>0$ to $j(t)$ evaluated at some specific values of $t.$  From assumption~\ref{assump-3} we know that $\beta$ is related to the lower bound in the forward limit ($t=0$) that can be taken into account in eq.~\eqref{bound-regge-traject-1} by including $a=1$ which corresponds to $\Delta=0$ and $C_0(\Delta=0)=0.$ Therefore, we have $\beta=-j(0).$ 
Moreover, the amplitude~\eqref{reggebehave} is a polynomially growing function of $s$ for $j(t)>0.$ Thus, from assumption~\ref{assump-2} we have $N=\max\limits_{z\in \partial\mathcal{D}_R} [{\rm Re}\,j(t(z))]$. In what follows, to keep the notation simple we just write $N$ instead of its explicit expression.

We can recast the inequality~\eqref{bound-regge-traject-1} in a more useful form by noticing that for sufficiently large values of $\Delta$ the function $\tilde{f}(\Delta)$ is smaller than its leading order in the $\Delta\gg 1$ expansion, i.e., we have
\begin{equation}
\tilde{f}(\Delta)\leq 2[N-j(0)]\sqrt{2\Delta}\,,\label{ineq-j-delta}
\end{equation}
where $N-j(0)=N+\beta>0$ by construction. More precisely, eq.~\eqref{ineq-j-delta}  holds true as long as $\Delta\geq \Delta_*,$ where $\Delta_*>0$ is defined as the solution of the equation $\tilde{f}(\Delta_*)=2[N-j(0)]\sqrt{2\Delta_*}.$ 

Therefore, by using~\eqref{ineq-j-delta} we obtain the following bound on the Regge trajectory:
\begin{equation}
\boxed{\max_{8m^2\Delta-4p_s^2\leq t\leq -8m^2\Delta} j(t)\geq -2[N-j(0)]\sqrt{2\Delta}}\quad \text{for}\quad \Delta\geq \Delta_*\,,\label{bound-regge-traject-3}
\end{equation}
which is non-trivial only for scenarios in which $j(t)<0$ for $8m^2\Delta-4p_s^2\leq t\leq -8m^2\Delta.$

\paragraph{Remarks.} Let us now make the following observations.

\begin{itemize}

\item  An approximate analytic expression for $\Delta_*$ can be found in the limit $\Delta\gg 1$ which also implies $t\ll -m^2.$ In such a regime we can write
\begin{equation}
\tilde{f}(\Delta)=2[N-j(0)]-\frac{1}{2}[3N-j(0)]+\frac{7}{8}[N-j(0)]\frac{1}{\sqrt{2\Delta}}+\mathcal{O}(1/\Delta)\,,
\end{equation}
and solve the equation $\tilde{f}(\Delta_*)=2[N-j(0)]\sqrt{2\Delta_*}$ up to this order in the expansion. By doing so, we obtain
\begin{equation}
\Delta_*\simeq \frac{49}{32}\left[\frac{N-j(0)}{3N-j(0)}\right]^2\,.
\end{equation}

\item In the regime $s\gg -t\geq 8m^2\Delta\geq 8m^2\Delta_*$ the bound~\eqref{bound-regge-traject-3} is consistent with the original Cerulus-Martin bound~\eqref{C-M-bound}. Indeed, since $s> 8m^2\Delta$ we have $\sqrt{s}>\sqrt{8m^2\Delta}$ and
\begin{equation}
\max_{8m^2\Delta-4p_s^2\leq t\leq -8m^2\Delta} j(t)> -[N-j(0)]\sqrt{s}\,,\label{bound-regge-traject-s}
\end{equation}
which implies
\begin{align}
 \max\limits_{8m^2\Delta-4p_s^2\leq t\leq -8m^2\Delta} \left(\frac{s}{s_0}\right)^{j(t)}>  \left(\frac{s}{s_0}\right)^{-2[N-j(0)]\sqrt{2\Delta}}> e^{-[N-j(0)]\sqrt{s/m^2}\ln (s/s_0)}\,. \label{regge-traj-MC}
\end{align}

\end{itemize}

\subsubsection{Case of monotonically increasing Regge trajectory for $t<0$} \label{sec:regge-traj-monot}

We now apply the above result to the case of Regge trajectories that are monotonically increasing for negative $t,$ i.e., we assume\footnote{Strictly speaking, we only need \eqref{reggemax} to obtain \eqref{bound-regge-traject-posit-der}. Monotonicity \eqref{posit-der} is a sufficient condition for \eqref{reggemax}.}
\begin{align}
\frac{{\rm d} j(t)}{{\rm d}t}\geq 0
\quad \text{for}\quad  t<0\,.
\label{posit-der}
\end{align}
If eq.~\eqref{posit-der} is valid,   we have 
\begin{equation}
\max\limits_{8m^2\Delta-4p_s^2\leq t\leq -8m^2\Delta} j(t)=j(-8m^2\Delta),\label{reggemax}
\end{equation}
and thus we can recast the bound~\eqref{bound-regge-traject-3} in the following more suggestive form:
\begin{equation}
\boxed{j(t)\geq -[N-j(0)]\sqrt{-t/m^2}\,}\quad \text{for}\quad t\leq -8m^2\Delta_*\,,\label{bound-regge-traject-posit-der}
\end{equation}
where we recall that $N-j(0)>0$ by construction and $\Delta_*>0$ was defined by the equation $\tilde{f}(\Delta_*)=2[N-j(0)]\sqrt{2\Delta_*}.$

The inequality~\eqref{bound-regge-traject-posit-der} implies that in the case of local theories with $\alpha=0$ a negative Regge trajectory $j(t)$ cannot decrease faster than $\sqrt{-t}$ for momentum transfer squared in the range $-\infty< t\leq -8m^2\Delta_*.$  For example, linear Regge trajectories for negative $t$ are not allowed. On the other hand, the lower bound~\eqref{bound-regge-traject-posit-der} does not constrain the Regge trajectory in the regime $t> -8m^2 \Delta_*;$ in particular, it does not say anything about the Regge trajectory for positive $t.$ 

Note that, the Veneziano amplitude~\cite{Veneziano:1968yb} in perturbative string theory violates both the bound~\eqref{bound-regge-traject-posit-der} and the Cerulus-Martin bound. Indeed, the Regge trajectory of the Veneziano amplitude is monotonically increasing and linear in the regime $-s\ll t\leq 0,$ thus it violates the lower bound~\eqref{bound-regge-traject-posit-der}.  Moreover, in the hard-scattering regime  the Veneziano amplitude behaves as $\sim e^{-\xi(\cos\theta)\,s},$ thus signaling a violation of~\eqref{C-M-bound}. See sec.~\ref{sec:examp-string} for more discussions on this point and for the explicit expression of $\xi(\cos \theta)$.

\section{Applications}\label{sec:examples}

We are now ready to consider applications of our results in various contexts. Since the newly derived bounds on elastic scattering amplitudes concern the high-energy regime, they can be used as powerful probes of non-locality and as tools to constrain UV physics.

Given a scenario in which the high-energy behavior of an elastic scattering amplitude is known, then we can use the newly derived bounds to check whether the starting assumptions are satisfied. In particular, the degree of (non-)localizability of the underlining UV theory can be constrained as follows. Assume that in a certain framework/scenario the high-energy behavior of an elastic scattering amplitude in the hard-scattering regime goes as  
\begin{equation}
|\scat(s,z)|\sim e^{-c\, s^\gamma}\,,\label{examp-UV}
\end{equation}
where $c>0$ and $\gamma>0$ are independent of $s.$ Then, to be consistent with the lower bound in eq.~\eqref{new-bound} we must impose the condition
\begin{equation}
\alpha\geq \gamma-\frac{1}{2}\quad (\text{hard-scattering regime})\,,
\label{constraint-hard}
\end{equation}
which represents a non-trivial constraint on the degree of non-locality of the UV theory that predicts the high-energy behavior in~\eqref{examp-UV}. Indeed, if $\gamma \leq 1/2$ the Cerulus-Martin bound and the property of polynomial boundedness can still be respected. On the other hand, if $\gamma>1/2$ we must have $\alpha>0$ which implies that the underlining theory is either strictly localizable, quasi-local or non-localizable (see eq.~\eqref{eq:classific}).

A similar discussion can be made in the Regge regime. Assume that the high-energy behavior of an elastic scattering amplitude in the Regge limit behaves as~\eqref{examp-UV}, then to be consistent with the lower bound in eq.~\eqref{new-bound-regge} we must impose the condition
\begin{equation}
\alpha\geq \gamma\quad (\text{Regge regime})\,.
\label{constraint-regge}
\end{equation}
It follows that the lower bound in the Regge limit~\eqref{new-bound-regge} can be more restrictive than the one in the hard-scattering regime~\eqref{new-bound}, i.e., eq.~\eqref{constraint-regge} is stronger than eq.~\eqref{constraint-hard}. Of course, the constraint~\eqref{constraint-regge} can only be useful if the elastic scattering amplitude is exponentially suppressed in the Regge regime, which however is usually not the case.

Hence, the newly derived bounds turn out to be very powerful tools to constrain  approaches to UV completion such as string theory~\cite{Gross:1987kza,Gross:1987ar,Mende:1989wt} and classicalization~\cite{Dvali:2010jz,Keltner:2015xda}. In addition, our results can be important to deepen our understanding of gravitational processes at high energies~\cite{Giddings:2007qq,Giddings:2009gj}, e.g., black-hole formation.
In what follows we are going to discuss applications of our results to specific theories/scenarios. While in sec.~\ref{subsec:candidate} we have shown examples of non-localizable theories by mainly looking at the spectral densities, in the next subsection we will analyse the behavior of the scattering amplitudes and get information about the degree of non-localizability (i.e., the value of $\alpha$) by using our lower bounds as probes.

\subsection{Gravity and black-hole formation}\label{sec:examp-BH-GR}

As discussed in sec.~\ref{sec:cand-gravity}, in Einstein's gravity one expects that the high-energy  behavior of an elastic scattering amplitude is characterized by an exponential suppression caused by BH production. In the hard-scattering regime (i.e., fixed angle) and in $D$ dimensions we have~\cite{Arkani-Hamed:2007ryv,Dvali:2014ila}
\begin{equation}
\sqrt{s}\gg M_p\quad \Rightarrow\quad |\scat(s,z)|\sim e^{-c\, (s/M_p^2)^{\frac{1}{2}\frac{D-2}{D-3}}}\,,\label{UV-gravity}
\end{equation}
where $c$ is some positive parameter independent of $s.$ By comparing with~\eqref{C-M-bound}, we can immediately notice that the Cerulus-Martin bound is always violated for any $D>3.$ Thus, it is very important to understand which of the assumptions in the derivation by Cerulus and Martin is violated.

We expect analyticity (assumption~\ref{assump-1}) and the lower bound on the forward limit in eq.~\eqref{bound-forward} (assumption~\ref{assump-3}) to be satisfied~\cite{Giddings:2007qq}. Whereas, we already know from sec.~\ref{sec:cand-gravity} that polynomial boundedness (assumption~\ref{assump-2} with $\alpha=0$) should be violated in gravitational theories. In addition, because of the massless nature of the graviton the assumption of a finite mass gap is also not satisfied. However, the high-energy behavior in eq.~\eqref{UV-gravity} should be insensitive to the presence of massless poles. In fact, as shown in~\cite{Tourkine:2023xtu} at least for $\alpha=0$ the derivation of~\eqref{new-bound} can be generalized to the case of gapless theories by suitably changing the definitions of domains in the complex $z$-plane (see also footnote~\ref{foot-gapless}).
Therefore, it is reasonable to conclude that the violation of the Cerulus-Martin bound by~\eqref{UV-gravity} is due to the lack of polynomial boundedness or, in other words, to a violation of locality in gravity; see also~\cite{Giddings:2007bw} for a similar reasoning. This means that the consistency of eq.~\eqref{UV-gravity} must be checked by making a comparison with the more general lower bound~\eqref{new-bound} derived under the assumption of exponential boundedness (assumption~\ref{assump-2}).

In the hard-scattering regime, we have to impose the consistency condition~\eqref{constraint-hard} with $\gamma=\frac{1}{2}\frac{D-2}{D-3},$ thus we obtain the following constraint on the degree of non-locality in gravity with $D>3$:
\begin{equation}
\alpha\geq \frac{1}{2(D-3)}\,.\label{gravity-constraint}
\end{equation}

In $D=4,$ we have $\alpha\geq 1/2,$ which means that Einstein's theory is either quasi-local or non-localizable. However, in higher dimensions $D>4$, the bound \eqref{gravity-constraint} also admits the value $\alpha<1/2$ and the strictly localizable case is not excluded. By contrast, the discussion in sec.~\ref{sec:cand-gravity} based on the BH entropy argument suggest that Einstein's GR in asymptotically flat spacetime is always non-localizable with $\alpha>1/2$ for any dimensions $D>3$. This discrepancy between the conclusion based on our bound \eqref{constraint-hard} and the BH entropy argument is due to the $1/2$ term on the RHS of \eqref{constraint-hard}. It would be then interesting to study if we can actually refine our current lower bound on the fixed angle scattering to replace \eqref{constraint-hard} by the stronger constraint $\alpha \geq \gamma$. 
We leave this task for future work.

\subsection{Classicalization proposal}\label{sec:examp-classicaliz}

In the more general case of the classicalization proposal, we can analyse the high-energy behavior of the S-matrix for a generic interaction potential of the form $\partial^{2k}\phi^n.$ As mentioned in sec.~\ref{sec:cand-classicaliz}, in this case it was argued that the high-energy behavior of the amplitude in the hard-scattering regime and in $D=4$ is suppressed as~\cite{Dvali:2010jz}
\begin{equation}
\sqrt{s}\gg \Lambda \quad \Rightarrow\quad |\scat(s,z)|\sim e^{-c\, (s/\Lambda^2)^{\frac{n+2k-4}{n+4k-6}}}\,,
\label{UV-classicaliz}
\end{equation}
where $\Lambda$ is a cutoff energy scale at which the perturbative EFT description breaks down. In this case the Cerulus-Martin bound is always violated for any $n\geq 3,$ and it is still reasonable to assume that the reason lies in the lack of polynomial boundedness. 

Therefore, we expect eq.~\eqref{UV-classicaliz} to be consistent with the more general bound~\eqref{new-bound}. By imposing the consistency condition~\eqref{constraint-hard}, we obtain the following non-locality constraint:
\begin{equation}
\alpha\geq \frac{1}{2}\frac{n-2}{n+4k-6}\,.
\label{constraint-classicaliz}
\end{equation}
Note that for $k=1$ we recover the gravitational case $\alpha\geq 1/2$ in eq.~\eqref{gravity-constraint} with $D=4$ dimensions. For other values of $k$ the constraint~\eqref{constraint-classicaliz} is not very strong because it would also allow for values  $\alpha<1/2.$ However, from the discussion in sec.~\ref{sec:cand-classicaliz} we expect that $\alpha> 1/2$ for $n\geq 3.$ As we discussed in sec.~\ref{sec:examp-BH-GR}, this discrepancy is due to the $1/2$  term on the RHS of~\eqref{constraint-hard}.

\subsection{Galileon}\label{sec:examp-galileon}

In Ref.~\cite{Keltner:2015xda} it was argued that at the non-perturbative level the Galileon violates the Cerulus-Martin bound because the fixed-angle elastic amplitude should behave as
\begin{equation}
\sqrt{s}\gg \Lambda \quad \Rightarrow\quad |\scat(s,z)|\sim e^{-c\, (s/\Lambda^2)^{\frac{3}{5}}}\,.
\label{UV-galileon}
\end{equation}
This conjectured behavior was hinted by looking at the behavior of the non-perturbative spectral density which was shown to be exponentially growing with $\alpha=3/5$~\cite{Keltner:2015xda}, as mentioned in sec.~\ref{sec:cand-galil} and in agreement with a UV completion by classicalization.

Hence, the consistency constraint on the exponential behavior of the Galileon model is a special case of eq.~\eqref{constraint-classicaliz} obtained for the classicalization proposal. We have
\begin{equation}
\alpha\geq \frac{1}{10}\,,
\label{constraint-galileon}
\end{equation}
which turns out to be a  weak constraint on the degree of non-locality because it would also allow for strict localizability. This is again due to the $1/2$ term on the RHS of \eqref{constraint-hard}. In fact, as discussed in sec.~\ref{sec:cand-galil}  we expect the Galileon field to belong to the class of non-localizable QFT with $\alpha=3/5.$

\subsection{Perturbative string theory}\label{sec:examp-string}

Let us now analyse scattering amplitudes in perturbative string theory and ask whether the Cerulus-Martin bound and the polynomial boundedness condition are violated. 
Let us consider the tree-level elastic scattering amplitude between tachyons in open string theory of the form~\cite{Polchinski:1998rq}
\begin{equation}
\mathcal{M}(s,t)=\mathcal{A}(s,t)+\mathcal{A}(s,u)+\mathcal{A}(t,u)\,,
\label{venez-ampl-tot}
\end{equation}
where
\begin{equation}
\mathcal{A}(s,t)=\frac{\Gamma(-j(s))\Gamma(-j(t))}{\Gamma(-j(s)-j(t))}\,,
\label{venez-ampl}
\end{equation}
is the Veneziano amplitude~\cite{Veneziano:1968yb}; similar formulas can be written for $\mathcal{A}(s,u)$ and $\mathcal{A}(t,u).$ $\Gamma(x)$ is the Euler gamma function, and the argument $x=-j(s)$  is linear in $s$, 
\begin{equation}
j(s)=j_0+\alpha^\prime s\,,
\end{equation}
where $j_0\equiv j(0)$ and $\alpha^\prime$ are positive constants. 
Remember that $u=-s-t+4m^2$ where, in this case, $m^2=-1/\alpha^\prime$ is the tachyon mass. The Veneziano amplitude has an infinite number of poles located on the real axis, e.g., the first term~\eqref{venez-ampl} gives $s,t$-poles at
\begin{equation}
s,t=\frac{n-j_0}{\alpha^\prime}\,,\quad n=0,1,2,\dots\,.
\label{venez-poles}
\end{equation}
One usually has $j_0=1,$ thus both massless (for $n=1$) and massive particles are included in the physical spectrum. 

Now we study the high-energy behavior of the amplitude~\eqref{venez-ampl-tot} in both hard-scattering (fixed $z=\cos\theta$) and Regge (fixed $t$) regimes. It is sufficient to focus on $\mathcal{A}(s,t)$ because similar results can be found for the other two terms. For fixed angle we have:
\begin{equation}
\text{hard-scattering regime:}\quad \mathcal{A}(s,t(z))\sim g(s,z)\,e^{-\xi(z)\,j(s)}\sim e^{-\xi(z)\,\alpha^\prime\,s}\,,
\label{venez-hard}
\end{equation}
where $g(s,z)$ is subdominant in the large $s$ limit and $ \xi(z)=\frac{1-z}{2}\log(\frac{2}{1-z})+\frac{1+z}{2}\log(\frac{2}{1+z})$ is positive for physical scattering angles. 
For fixed momentum transfer, instead, we obtain the following polynomial behavior:
\begin{equation}
\text{Regge regime:}\quad \mathcal{A}(s,t)\sim B(s,t)\,s^{j(t)}\,,
\label{venez-regge}
\end{equation}
where $B(s,t)$ is subdominant in the large $s$ limit, and $j(t)=j_0+\alpha^\prime t$ is the Regge trajectory. 

As it is well known, eq.~\eqref{venez-hard} violates the Cerulus-Martin bound~\eqref{C-M-bound}. In addition, the linear Regge trajectory in eq.~\eqref{venez-regge} violates the bound in~\eqref{bound-regge-traject-posit-der} that we derived for negative $t$ in the case $\alpha=0$. 
As it is pointed out in \cite{Gross:1987kza}, the reason behind the violation of the Cerulus-Martin bound would lie in the lack of polynomial boundedness. 
This is because, even if we tune the parameters $j_0$ and $\alpha^\prime$ so that no massless pole appears (i.e., $j_0-n\neq 0$) we will still have the same high-energy behaviors in the two regimes. Moreover, both assumptions~\ref{assump-1} and~\ref{assump-3} can be satisfied by the amplitude~\eqref{venez-ampl}.
Indeed, we can verify that there exists some value $z_*$ for which the amplitude $\mathcal{M}(s,t(z))$ grows exponentially as $e^{+{\rm Re}[\xi(z_*)]\,s}$ in the limit $s\rightarrow \infty,$ where ${\rm Re}[\xi(z_*)]>0.$ This means that (perturbative) string theory is characterized by features that cannot be described in the context of local QFT and falls into the class $\alpha=1$ according to our parameterization.

A similar analysis can be performed in the case of closed string theory for the tree-level closed string exchange, e.g., for the Virasoro-Shapiro amplitude~\cite{Virasoro:1969me,Polchinski:1998rq}. Also in this case the high-energy behavior of the scattering amplitude is not consistent with polynomial boundedness. In fact, the hard-scattering and Regge behaviors are the same as in the case of Veneziano amplitude.

It is worth mentioning that the violation of the Cerulus-Martin bound in perturbative string theory was initially discussed in Ref.~\cite{Gross:1987kza,Gross:1987ar}. Here, by using the  more general lower bound~\eqref{new-bound} we can understand the reason behind such a violation more concretely and that it does not imply a pathology because the high-energy behavior of the Veneziano amplitude must be checked to be consistent with a different lower bound, i.e., the one in eq.~\eqref{new-bound} with $\alpha=1,$ which is indeed satisfied by~\eqref{venez-hard}. 

Despite the non-local nature of scattering amplitudes evaluated perturbatively, Ref.~\cite{Mende:1989wt} claims that the Cerulus-Martin bound is restored in the non-perturbative amplitude obtained after the resummation of divergent perturbative expansion. This result is interpreted as an improvement of the locality of strings~\cite{Mende:1989wt}. 
It may be then interesting to investigate if the polynomial boundedness \eqref{assump-exp} with $\alpha=0$ is also recovered nonperturbatively. For this to happen, the linearity of $j(t)$ would be modified at least for complex $t$ with large modulus because of the following reason. The kinematics $(s,t(z))$ with $s\gg m^2$ is well inside the Regge regime when $z\in\partial \mathcal D_R$ is in the vicinity of the point $z=\rho$. Except for the point $z=\rho$, $t$ becomes complex and both its real part and imaginary part grow linearly as $s\in\mathbb R$ increases. Hence, the high-energy behavior of $\scat(s,t(z))$ within such kinematics can be captured by the Regge behavior $\scat\sim s^{j(t)}$ with complex $t$ with large modulus. In particular, if $j(t)$ is linear even for complex $t$, it violates the polynomial boundedness. We leave the 
study of $j(t)$ for complex $t$ with large modulus as a future work.

\subsection{Infinite-derivative QFTs}\label{sec:examp-inf-der-QFT}

To conclude the list of applications, we now consider a non-local QFT in which the Lagrangian contains infinite-order differential operators. In particular, we focus on the following type of Lagrangian~\cite{Tomboulis:2015gfa,Buoninfante:2018mre}
\begin{equation}
\mathcal{L}=\frac{1}{2}\phi\, e^{(-\Box/M^2)^n}(-\Box+m^2)\,\phi -\frac{\lambda}{3!}\phi^3\,,\label{exp-non-local-phi^3}
\end{equation}
where $n$ is a positive integer. For these models the corresponding S-matrix respects unitarity (i.e., the optical theorem) at any loop-order in perturbation~\cite{Pius:2016jsl,Briscese:2018oyx,Chin:2018puw,Koshelev:2021orf,Buoninfante:2022krn}. 
It is also worth to mention that differential operators such as $e^{c\,\Box}$  are typical of string field theory~\cite{Witten:1985cc,Eliezer:1989cr,Pius:2016jsl} and p-adic string~\cite{Freund:1987kt,Brekke:1988dg}.

We are interested in the behavior of the $2\rightarrow2$ elastic scattering amplitude. At  tree level we have
\begin{equation}
\mathcal{M}^{(n)}(s,t)= \lambda^2 \left(\frac{e^{-(-s)^n/M^{2n}}}{s-m^2}+\frac{e^{-(-t)^n/M^{2n}}}{t-m^2}+\frac{e^{-(-u)^n/M^{2n}}}{u-m^2}\right)\,.
\label{tree-amplitude-n}
\end{equation}
Let us note that infinite-derivative QFTs with odd powers $n$ can be pathological because the scattering amplitude grows exponentially in $s$ in the Regge limit ($s\to\infty$ with fixed $t$) even when $t<0, s\in \mathbb{R}.$ In the case of string field theory similar exponential operators with $n=1$ are present but the form of the interaction vertices is sufficiently complicated such that no pathological behavior arises~\cite{Pius:2016jsl}.

As an example, we explicitly analyse the case  $n=2$ in which the perturbation theory is well-defined at high energy for physical values of the momenta.  
In this case, the tree-level amplitude~\eqref{tree-amplitude-n} reduces to
\begin{equation}
\mathcal{M}^{(2)}(s,t)= \lambda^2 \left(\frac{e^{-s^2/M^{4}}}{s-m^2}+\frac{e^{-t^2/M^{4}}}{t-m^2}+\frac{e^{-u^2/M^{4}}}{u-m^2}\right)\,.
\label{tree-amplitude-2}
\end{equation}
Obviously, the high-energy behavior of this amplitude for fixed angle is in contradiction with the Cerulus-Martin bound~\eqref{C-M-bound}. Indeed, in the hard-scattering regime we need to take the limit $s\rightarrow \infty$  with $z=\cos\theta=\text{fixed},$ thus we obtain 
\begin{equation}
\text{hard-scattering regime:}\quad \mathcal{M}^{(2)}(s,t(z))\sim \lambda^2 \left(\frac{e^{-s^2/M^4}}{s}-2\frac{e^{-s^2(1-z)^2/4M^4}}{s(1-z)}-2\frac{e^{-s^2(1+z)^2/4M^4}}{s(1+z)}\right)\,,
\end{equation}
which decays faster than $e^{-c\,\sqrt{s}\log s}.$

We now want to check whether all the assumptions~\ref{assump-1},~\ref{assump-2} and~\ref{assump-3} are satisfied and, in particular, that polynomial boundedness is violated. First of all, the analytic structure of the amplitude is the same as the one in the standard local case ($M\rightarrow \infty$) besides the presence of essential singularities at infinity. In the forward limit ($t=0$) and at high energy the amplitude tends to a constant, thus respecting eq.~\eqref{bound-forward}. Moreover, we can check that the amplitude is exponentially bounded for $z\in\mathbb{C}$ and large $s\in \mathbb{R}.$ Since there exists some value $z_*$ such that ${\rm Re}[(1+z_*)^2]<0,$ the amplitude~\eqref{tree-amplitude-2} can diverge in the complex $z$-plane for large $s$ as
\begin{equation}
\mathcal{M}^{(2)}(s,t(z_*))\sim  \lambda^2\frac{e^{+c\,s^2|{\rm Re}[(1+z_*)^2]|/M^4}}{s}\,,
\end{equation}
where $c$ is some positive constant. This behavior suggests that the infinite-derivative theory~\eqref{exp-non-local-phi^3} with $n=2$ is  non-localizable  with degree of non-locality given by $\alpha=n=2.$ 

Hence, we can conclude that the violation of the Cerulus-Martin bound is not a pathology of the model~\eqref{exp-non-local-phi^3} because polynomial boundedness is not satisfied.
In fact, the amplitude~\eqref{tree-amplitude-2} must satisfy a more general lower bound derived under the assumption of exponential boundedness and, indeed, we can easily check that the high-energy behavior for fixed angle is consistent with the lower bound~\eqref{new-bound} with $\alpha=2$ derived in the hard-scattering regime.

\section{Summary \& conclusions}\label{sec:discuss}

In this work we derived new rigorous lower bounds on elastic scattering amplitudes in two different kinematic regimes: hard-scattering and Regge limits. The working assumptions are (see also sec.~\ref{subsec:assumptions}): (\ref{assump-1})~analyticity in a certain subdomain $\mathcal{D}_R$ of the complex $z$-plane (where $z=\cos\theta$), (\ref{assump-2})~exponential boundedness of the amplitute in $\mathcal{D}_R$, (\ref{assump-3})~forward amplitude bounded from below by inverse powers of the center-of-mass energy, and (4)~a finite mass gap. The assumption~\ref{assump-2} on the exponential growth of the amplitude is physically related to the non-local nature of the corresponding underlining theory. In particular, in our study we proposed to define and parametrize non-locality in terms of the S-matrix, and our parameterization in eq.~\eqref{eq:basicassump} or~\eqref{assump-exp} is motivated by Jaffe's criterion~\eqref{eq:classific} for the Wightman functions.

To clearly summarize our results we now write down again the newly derived lower bounds.
In the hard-scattering limit we obtained (in eq.~\eqref{new-bound})
\begin{equation}
\max_{-a\leq \cos\theta\leq a}\left|\mathcal{M}(s,\cos\theta)\right|\geq \mathcal{N}(s)\, e^{-f(a)\,\sqrt{s}\,\log (s/s_0)} e^{-g(a)\,s^{\alpha+\frac{1}{2}}}\,,
\end{equation}
where $\mathcal{N}(s)$ is a positive function of $s$ that is subdominant in the $s\rightarrow\infty$ limit, and $f(a)$ and $g(a)$ are some positive functions of $a\in (0,1)$. In the case of polynomial boundedness ($\alpha=0$) we consistently recovered the Cerulus-Martin lower bound~\cite{Cerulus:1964cjb}, i.e., eq.~\eqref{C-M-bound}. Whereas, in the case of exponential boundedness ($\alpha>0$) the scattering amplitude is more suppressed in the high-energy limit and thus the violation of the Cerulus-Martin bound is allowed. 

In the Regge limit we obtained (in eq.~\eqref{new-bound-regge})
\begin{equation}
\max_{8m^2\Delta-4p_s^2\leq t\leq -8m^2 \Delta}\left|\mathcal{M}(s,t)\right|\geq h(\Delta) \, e^{-\tilde{f}(\Delta) \log(s/s_0)-\tilde{g}(\Delta)s^\alpha}\,, 
\end{equation}
where $h(\Delta),$ $\tilde{f}(\Delta)$ and $\tilde{g}(\Delta)$ are positive functions of $0<\Delta\leq p_s^2/4m^2,$ see eq.~\eqref{func-def-regge}. This lower bound turns out to be a new result in both cases of polynomial and exponential boundedness. In particular, in the former case ($\alpha=0$) the lower bound is given by inverse powers of $s,$ whereas in the latter case ($\alpha>0$) an exponential suppression of the amplitude is allowed also in the Regge limit. Note that, the bound in the Regge limit is stronger than the one in the hard-scattering regime due to a missing factor $\sqrt{s}$ in the exponent. This means that, in principle, scattering amplitudes at large angles are allowed to be more suppressed as compared to the ones at small angles. 

It should be emphasized that the above two inequalities are not bounds at some specific $\theta$ or $t$ because any amplitude can in principle vanish at given $\theta$ or $t$. Indeed, they are lower bounds on the maximum of the modulus of the amplitude for some finite intervals of $\theta$ or $t$.

Furthermore, as a consequence of the bound in the Regge regime, we studied the implications for the Regge trajectory. Given a Regge behavior of the form $s^{j(t)},$ where $j(t)$ is the Regge trajectory, in eq.~\eqref{bound-regge-traject-3} we obtained a new bound on $j(t)$ for negative $t.$ In particular, for a monotonically increasing and negative Regge trajectory, in eq.~\eqref{bound-regge-traject-posit-der} we obtained the following suggestive bound:
\begin{equation}
j(t)\geq -[N-j(0)]\sqrt{-t/m^2}\,,
\end{equation}
for $j(t)<0$ and $t\leq -8m^2\Delta_*,$ where  $N-j(0)>0$ by construction and $\Delta_*>0$ was defined by the equation $\tilde{f}(\Delta_*)=2[N-j(0)]\sqrt{2\Delta_*};$ see sec.~\ref{sec:regge-traj-monot} for more details. Such a bound on the Regge trajectory is derived under the assumption of polynomial boundedness ($\alpha=0$) and is very important because it forbids linear behaviors for negative $t.$ This also provides a clear explanation why the linear behavior of the Regge trajectory for the Veneziano amplitude in perturbative string theory causes a violation of the Cerulus-Martin bound, as we discussed in detail in sec.~\ref{sec:examp-string}.

Our results provide rigorous and quite general lower bounds on elastic scattering amplitudes. Moreover, they can be useful tools to constrain and/or gain 
information on the underlining UV theories. In sec.~\ref{sec:examples} we investigated several applications of the newly derived lower bounds.  By analysing the high-energy behavior of scattering amplitudes and BH formation in gravity we used the lower bound in the hard-scattering regime to constrain the degree of non-locality of gravitational theories. 
In particular, we found a new non-locality bound given by $\alpha\geq 1/2$ in $D=4$ spacetime dimensions, which suggests that gravitational theories (such as Einstein's general relativity) 
may have fundamentally non-local characteristics.
In addition, we studied the high-energy behavior of scattering amplitudes in the classicalization proposal, the Galileon model, perturbative string theory, and infinite-derivative QFTs. By means of our lower bounds, we constrained their degree of non-locality. 

The non-locality bound turns out to be quite weak for the Galileon model, but it can be stronger in the more general case of classicalization for certain values of the number of fields and derivatives in the interaction potential. In perturbative string theory we obtained $\alpha\geq 1/2,$ implying that tree-level string amplitudes cannot be captured by a local QFT description.

Hence, the newly derived bounds can act as probes of non-locality in various contexts. This also means that they can be used as consistency checks: they must be respected in any future approach to UV completion which violates the polynomial boundedness condition. In this paper we mainly used the bound in the hard-scattering regime to constrain the degree of non-locality. As a future work, it would be interesting to consider scenarios in which the amplitude is exponentially suppressed even in the Regge limit. In such a case, we would be able to obtain a stronger constraint on the degree of non-locality of the underlining UV theory, as explained at the beginning of sec.~\ref{sec:examples}.

Finally, let us emphasize that strictly speaking the derivations of the lower bounds elaborated here are only valid in the presence of a finite mass gap and that all the scenarios considered as applications do not respect such an assumption. However, as explained in sec.~\ref{sec:examples}, it is reasonable to expect that the real reason why the Cerulus-Martin bound is violated is the lack of polynomial boundedness. In fact, as already mentioned in several places in the paper, e.g., in the footnote~\ref{foot-gapless}, one can try to generalize our derivations to the case of vanishing mass gap by suitably changing the definitions of the dumb-bell domains in the complex $z$-plane. For example, this can be done by changing the definition of the domain $D_E$ in the complex $z$-plane, i.e., by modifying assumption~\ref{assump-3} away from the forward limit as was recently done for the Cerulus-Martin bound (i.e., $\alpha=0$ and hard-scattering regime) in Ref.~\cite{Tourkine:2023xtu}. It would be interesting to explicitly generalize our derivations under the assumption of zero mass gap for any $\alpha\geq 0$ and in both cases of hard-scattering and Regge limits. We expect that the only changes in the formula of the lower bounds are different expressions for the functions $f(a),$ $g(a),$ $\tilde{f}(\Delta)$ and $\tilde{g}(\Delta)$ in eqs.~\eqref{new-bound} and~\eqref{new-bound-regge}. Our physical intuition supporting this expectation is that the scattering process at small impact parameter would be essentially insensitive to the presence of the mass gap.
A detailed study of the gapless case will be part of a future work.


\subsection*{Acknowledgements}

L.~B. is grateful to Anna Tokareva for her hospitality at Imperial College London during the final stages of this work and would like to thank The Royal Society for financial support. Nordita is supported in part by NordForsk.
J.T. is supported by IBS under the project code, IBS-R018-D1. 
M.Y. is supported by IBS under the project code, IBS-R018-D3, and by JSPS Grant-in-Aid for Scientific Research Number JP21H01080.


\appendix

\section{Analysis of the functions $C(x,a)$ and $C(x,\Delta)$} \label{app-C(x)}

In this appendix we wish to show more details on the functions\footnote{With an abuse of notation we are calling the functions $C(x,a)$ and $C(x,\Delta)$ with the same letter $C.$ However, it should be understood that the functional dependence on $(x,a)$ is different from the one on $(x,\Delta)$. Indeed, as already explained in the main text, the latter is defined in terms of the former as $C(x,\Delta)\equiv C(x,a)|_{a=1-x\Delta}$. Below we will make a similar abuse of notation when we will define $x_*(a)$ and $x_*(\Delta)$.}  $C(x,a)$ and $C(x,\Delta)$ whose behaviors were crucial for the derivations of the lower bounds in the hard-scattering (sec.~\ref{sec:lower-bound}) and in the Regge (sec.~\ref{sec:lower-bound-Regge}) limits; recall that $x=(2m/p_s)^2.$ Let us consider the two regimes separately.

\begin{figure}[t!]
	\centering
\subfloat[Subfigure 1 list of figures text][]{
		\includegraphics[scale=0.41]{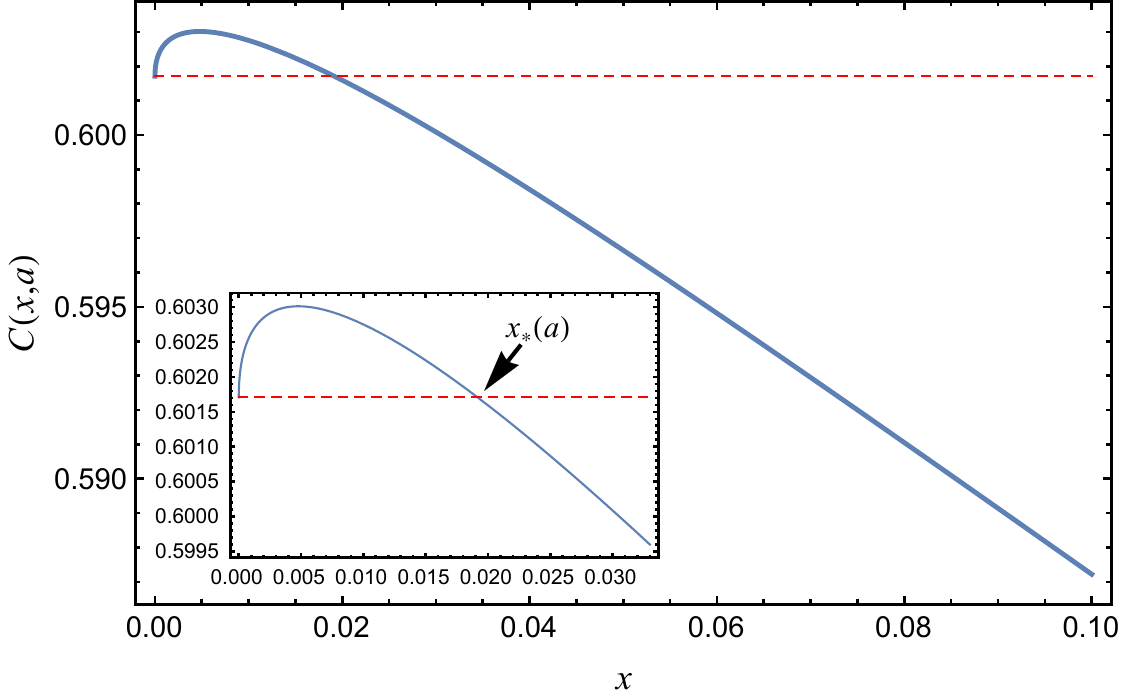}\label{figC}}\quad\,
	\subfloat[Subfigure 2 list of figures text][]{
		\includegraphics[scale=0.445]{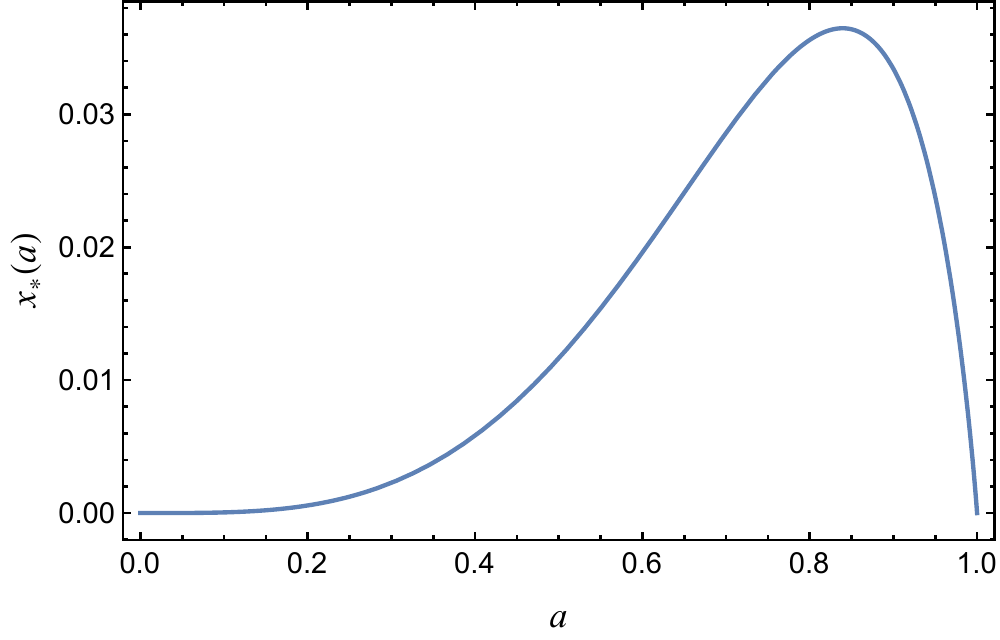}\label{fig-x_*(a)}}
 	\protect\caption{(a) Behavior of $C(x,a)$ in eq.~\eqref{function-C(x)} as a function of $x=(2m/p_s)^2$ (blue line). We can notice that for sufficiently small values of $x$ (sufficiently high energies) the inequality $C(x,a)\geq C_0(a)$ holds true, where $C_0(a)$ is given in eq.~\eqref{C0def}. We have set $a= 0.6,$ which implies $C_0(a)\simeq 0.601$ (red dashed line) and $x_*(a)\simeq 0.019$. (b) Plot of the analytic estimate~\eqref{x_*(a)-approx} for $x_*(a)$.}
\end{figure}

\paragraph{Hard-scattering regime.} In the limit of high energy and fixed-angle, we expanded the function~\eqref{function-C(x)} around $x=0$ (i.e., $s\rightarrow \infty$) and for fixed values of $a\in (0,1),$ and as a result we obtained $C(x,a)=C_0(a)+\mathcal{O}(\sqrt{x}),$ where $C_0(a)$ is given in eq.~\eqref{C0def}. A crucial step to derive the lower bound~\eqref{new-bound} is the use of the inequality
\begin{equation}
C(x,a)\geq C_0(a)\,.\label{ineq-hard}
\end{equation}
We can numerically show that~\eqref{ineq-hard} holds for all $x\leq x_*(a),$ where $x_*(a)\neq 0$ is defined as the solution of the equation  $C(x_*(a),a)=C_0(a).$ In Fig.~\ref{figC} we have shown the behavior of the function $C(x,a)$ as compared to $C_0(a)$ for $a=0.6.$ The figure suggests that the derivative $\der_x C(x,a)$ is positive at $x=0$ while it becomes negative at some point in the interval $0<x<x_*(a)$. 


This behavior of $C(x,a)$ can be understood analytically by working with an approximate analytic expression for $x_*(a)$. Let us expand $C(x,a)$ and its derivative as
\begin{eqnarray}
C(x,a)&=&C_0(a)+C_{1/2}(a)\sqrt{x}+C_1(a)x+\mathcal{O}(x^{3/2})\,,\label{C(x,a)-expans}
\\[2mm]
\frac{\der}{\der x} C(x,a)&=& \frac{1}{2}C_{1/2}(a)\frac{1}{\sqrt{x}}+C_1(a)+\mathcal{O}(\sqrt{x})\,,\label{derC_analytic}
\end{eqnarray}
where the expressions of the coefficients $C_{1/2}(a)$ and $C_1(a)$ are lengthy and we do not write them explicitly. Here, it is only important that $C_{1/2}(a)$ and $C_1(a)$ are always positive and negative for $0< a < 1$, respectively. Now we solve the equation $C(x_*(a),a)=C_0(a)$ by using~\eqref{C(x,a)-expans} and obtain
\begin{equation}
x_*(a)\simeq \left(-\frac{C_{1/2}(a)}{C_1(a)}\right)^2\,.\label{x_*(a)-approx}
\end{equation}
The analytic estimate~\eqref{x_*(a)-approx} is plotted in Fig.~\eqref{fig-x_*(a)}. 
We can notice that for $a=0.6$ we get $x_*(0.6)\simeq 0.019,$ which gives a good quantitative estimation of the intersection point between $C(x,a)$ and $C_0(a).$ The function $x_*(a)$ vanishes at $a=0$ and $a=1,$ i.e., $x_*(a=0)=0=x_*(a=1).$ Around these points, $x_*(a)$ is approximated as 
\begin{align}
    x_*(a)|_{a\sim0}
    \simeq
    \frac{9 a^4 \log ^2\left(\frac{4}{a}\right)}{\left(\log \left(\frac{4}{a}\right)+12\right)^2}
    \,,
    \qquad
     x_*(a)|_{a\sim1}
    \simeq
    \frac{8 (1-a)}{9}
    \,.
    \label{x_*(a)-approx_a1}
\end{align}
Moreover, from eq.~\eqref{derC_analytic} we can understand that $\der_x C(x,a)$ is positive for $0\leq x\lesssim \frac{x_*(a)}{4}$, while it becomes negative for $\frac{x_*(a)}{4}\lesssim x \leq x_*(a)$. This explains the behavior of $C(x,a)$ for $0\leq x\leq x_*(a)$ shown in Fig.~\ref{figC}.

\paragraph{Regge regime.} In the limit of high energy and fixed momentum transfer in sec.~\ref{sec:lower-bound-Regge} we considered an expansion of the function $C(x,\Delta)$ around $x=0$ and for fixed values of $\Delta.$ Also in this case, the behavior of the function at high energies was crucial for the derivation of the bound~\eqref{new-bound-regge}. 

First of all, by inspecting the structure of $C(x,\Delta)$ we can notice that it is real valued for $x\leq 1/\Delta,$ while it becomes complex valued for $x>1/\Delta.$  This happens because for $x>1/\Delta$ we get a logarithm with a negative argument, as it can be explicitly checked by taking $C(x,a)$ in eq.~\eqref{function-C(x)} and replacing $a\rightarrow 1-x\Delta.$ For completeness, here we show its full expression:
\begin{equation}
C(x,\Delta)=\frac{1}{\sqrt{x}}\frac{\log \left(\frac{2 \sqrt{(x+2)^2 \left(1-\frac{\left(-\sqrt{x \left(8 \Delta -4 \Delta ^2 x+x+4\right)}+x+2\right)^2}{4 (\Delta  x-1)^2}\right)}+2 x+4}{\sqrt{(x+2)^2 \left(\left(x-\sqrt{x (x+4)}+2\right)^2-\frac{\left(-\sqrt{x \left(8 \Delta -4 \Delta ^2 x+x+4\right)}+x+2\right)^2}{(\Delta  x-1)^2}\right)}+(x+2) \left(x-\sqrt{x (x+4)}+2\right)}\right)}{\log \left(\frac{(1-x\Delta) \left(2 \sqrt{(x+2)^2 \left(1-\frac{\left(-\sqrt{x \left(8 \Delta -4 \Delta ^2 x+x+4\right)}+x+2\right)^2}{4 (\Delta  x-1)^2}\right)}+2 x+4\right)}{(x+2) \left(-\sqrt{x \left(8 \Delta -4 \Delta ^2 x+x+4\right)}+x+2\right)}\right)}\,.
\label{C(x,delta)}
\end{equation}
This does not cause any problem in our analysis because we are interested in the parameter region $0\leq a\leq1$ that is equivalent to $x\leq 1/\Delta$. Thus, in what follows we assume the condition $x\leq 1/\Delta$.

\begin{figure}[t!]
	\centering
\subfloat[Subfigure 1 list of figures text][]{
		\includegraphics[scale=0.39]{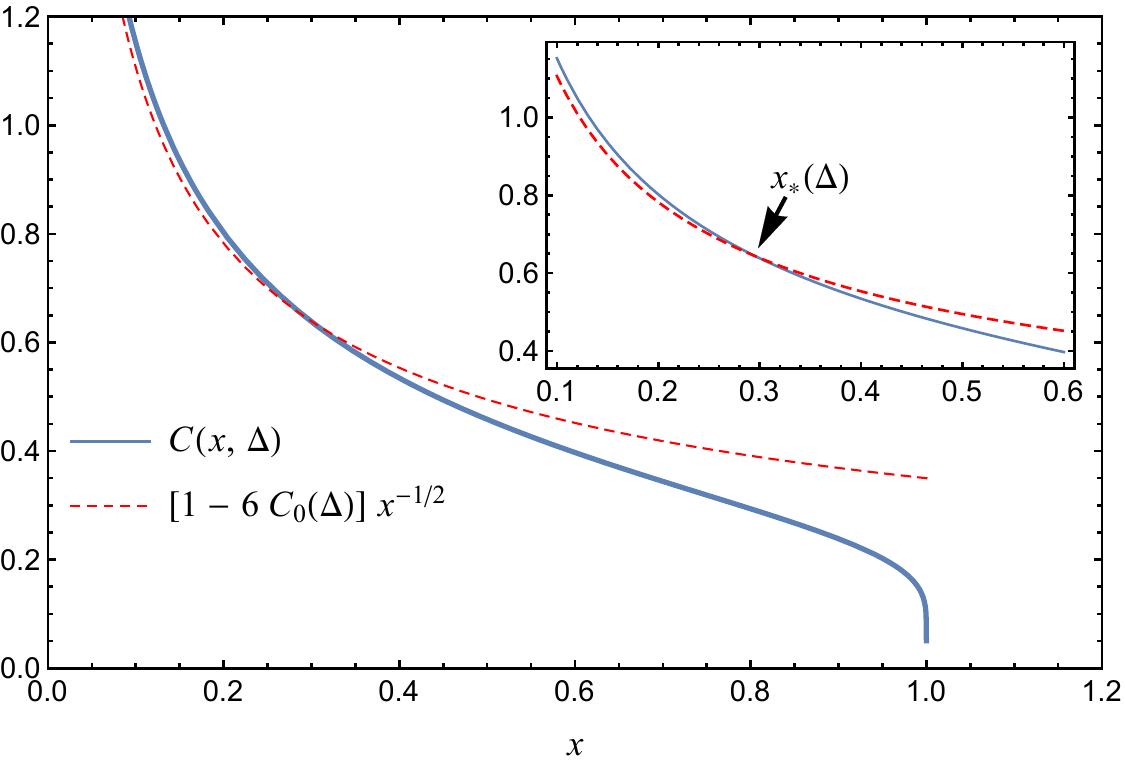}\label{figC-delta}}\quad\,
	\subfloat[Subfigure 2 list of figures text][]{
		\includegraphics[scale=0.377]{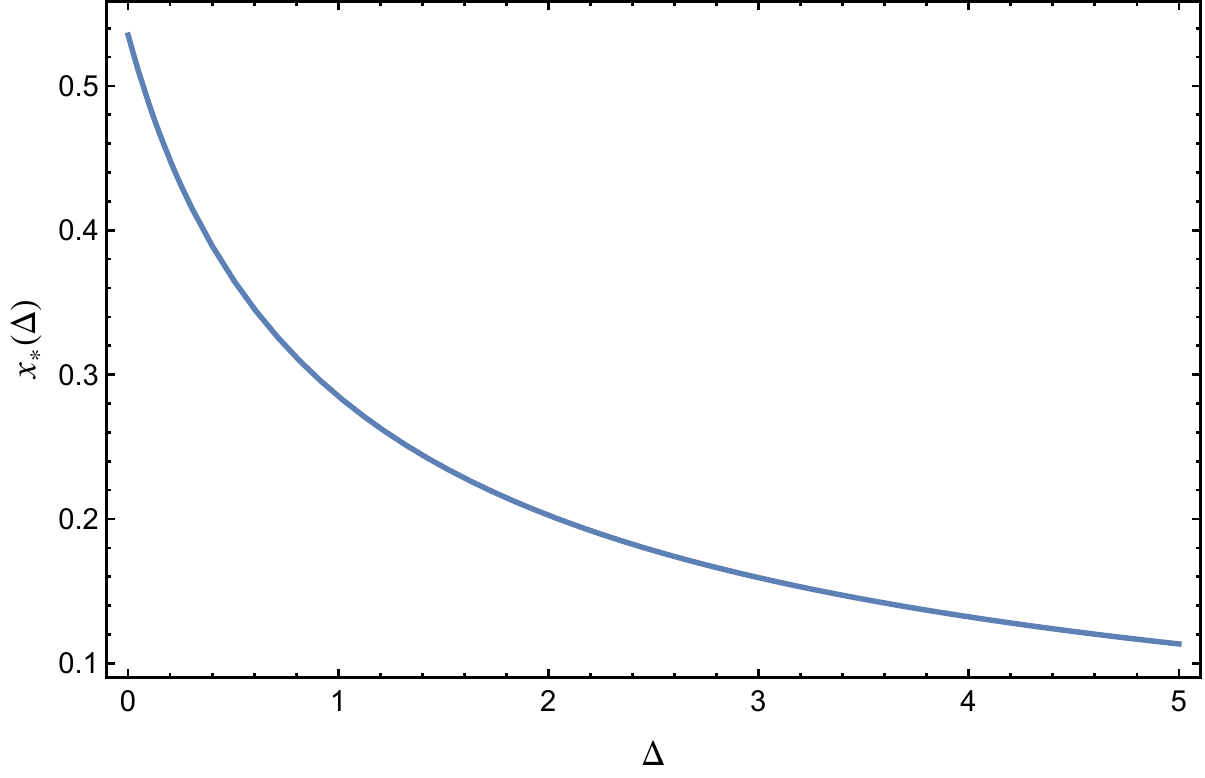}\label{fig-x_*(delta)}}
 	\protect\caption{(a) Behavior of $C(x,\Delta)$ in eq.~\eqref{C(x,delta)} as a function of $x=(2m/p_s)^2$ (blue line) as compared to $[1-6C_0(\Delta)]x^{-1/2}$  (red dashed line), where $C_0(\Delta)$ is given in eq.~\eqref{CDelta-def}. We can notice that for sufficiently small values of $x$ (i.e., sufficiently high energies) the inequality $C(x,\Delta)\geq [1-6C_0(\Delta)]x^{-1/2}$ holds true. We have set $\Delta= 1,$ which implies $C_0(\Delta)\simeq 0.1083$ and $x_*(\Delta)\simeq 0.2966$. Note that the curves stop at $x=1$ because the function $C(x,\Delta)$ is real only for $x\leq 1/\Delta;$  it becomes complex valued for $x>1/\Delta.$ (b) Plot of the approximate solution for $x_*(\Delta)$ in eq.~\eqref{x_*(delta)-approx}.}
\end{figure}

To prove~\eqref{new-bound-regge} it was crucial the use of the inequality 
\begin{equation}
C(x,\Delta)\geq [1-6C_0(\Delta)]\frac{1}{\sqrt{x}}\,.\label{ineq-regge}
\end{equation}
We can numerically show that \eqref{ineq-regge} holds for all $x\leq x_*(\Delta),$ where $x_*(\Delta)\neq 0$ is defined as the solution of the equation  $C(x_*(\Delta))=[1-6C_0(\Delta)](x_*(\Delta))^{-1/2}.$
In Fig.~\ref{figC-delta} we have shown the behavior of the function $C(x,\Delta)$ as compared to $[1-6C_0(\Delta)]x^{-1/2}$ for $\Delta=1.$ We can explicitly see that the sub-leading term $C_\text{sub}(x,\Delta)\equiv C(x,\Delta)-[1-6C_0(\Delta)]x^{-1/2}$ is non-negative and monotonically decreases in $x$ for $x\leq x_*(\Delta)$.


This behavior can be understood analytically by working with an approximate expression for $x_*(\Delta)$. Let us expand $C(x,\Delta)$ around $x=0$ as
\begin{align}
&C(x,\Delta)=C_{-1/2}(\Delta)\frac{1}{\sqrt{x}}+C_0(\Delta)+C_{1/2}(\Delta)\sqrt{x}+\mathcal{O}(x)\,,\label{C(x,delta)-expans}
\end{align}
where $C_{0}(\Delta)$ was defined in eq.~\eqref{CDelta-def}, whereas 
\begin{equation}
C_{-1/2}(\Delta)\equiv 1-6C_0(\Delta)\,,\,\quad C_{1/2}(\Delta)\equiv \frac{\Delta  \left(85-83 \sqrt{2 \Delta +1}\right)-2 \sqrt{2 \Delta +1}+2}{360 (2 \Delta +1)^{1/4} \left(\sqrt{2 \Delta +1}-1\right)^{3/2}}\,.
\end{equation}
From eq.~\eqref{C(x,delta)-expans} also follows that 
$C_\text{sub}$ can be approximated as $C_\text{sub}\simeq C_0(\Delta)+C_{1/2}(\Delta)\sqrt{x}$. This implies that  $C_\text{sub}$ is a monotonically decreasing function of $x$ because $C_{1/2}(\Delta)$ is negative for $\Delta>0$. 

The equation $C(x_*(\Delta),\Delta)=C_{-1/2}(\Delta)x_*^{-1/2}$ can be solved in the approximation~\eqref{C(x,delta)-expans}, and we obtain
\begin{equation}
x_*(\Delta)\simeq \left(-\frac{C_{0}(\Delta)}{C_{1/2}(\Delta)}\right)^2
=
\frac{7200}{6889 \Delta +6723 \sqrt{2 \Delta +1}+6725}
\,.\label{x_*(delta)-approx}
\end{equation}
We have shown the behavior of the analytic estimate \eqref{x_*(delta)-approx}  for $x_*(\Delta)$ in Fig.~\ref{fig-x_*(delta)}. $x_*(\Delta)$ is a monotonically decreasing function, it becomes maximum at $\Delta=0$ and vanishes in the limit $\Delta\rightarrow \infty.$ These qualitative features can be verified by looking at the intersection point of the full (non-expanded) $C(x,\Delta)$ and $C_{-1/2}(\Delta)x^{-1/2}$ by varying the values of $\Delta.$ In addition, we can also check that eq.~\eqref{x_*(delta)-approx} provides a good quantitative estimation: for $\Delta=1$ we have $x_*(\Delta=1)\simeq (-C_0(1)/C_{1/2}(1))^2\simeq 0.28$ (see Fig.~\ref{figC-delta}).


\bibliographystyle{utphys}
\bibliography{References}


\end{document}